%% file: v3.tex
\documentclass[a4paper]{IEEEtran}% ,double column

\include{preamble}

\begin{document}

%\title{Eigen-Inference Precoder: an EI Precoding Scheme Against Imperfect CSI for the Coarsely Quantized Massive MU-MIMO System}
%\title{Robust Hierarchical Precoding for Coarsely Quantized Massive MU-MIMO Systems with Imperfect Channel State Information}
\title{Eigen-Inference Precoding for Coarsely Quantized Massive MU-MIMO System with Imperfect CSI}

\author{Lei Chu, \IEEEmembership{Student Member, IEEE}, Robert Qiu, \IEEEmembership{Fellow, IEEE}, and Fei Wen \IEEEmembership{Member, IEEE},
\IEEEcompsocitemizethanks{
\IEEEcompsocthanksitem Lei Chu is with Department of Electrical Engineering, Research Center for Big Data Engineering Technology,
Shanghai Jiaotong University, Shanghai 200240, China. (leochu@sjtu.edu.cn)
\IEEEcompsocthanksitem Fei Wen is with Department of Electronic Engineering, Shanghai Jiaotong University,
Shanghai 200240, China. (wenfei@sjtu.edu.cn)
\IEEEcompsocthanksitem Dr. Qiu  is with the Department of Electrical and Computer Engineering\,
Tennessee Technological University, Cookeville, TN 38505 USA. Dr. Qiu is also with Department of Electrical Engineering,
Research Center for Big Data Engineering Technology, Shanghai Jiaotong University, Shanghai 200240, China. (e-mail: rqiu@tntech.edu). \protect
}
\thanks{Dr. Qiu's  work  is partially supported  by  N.S.F. of China under Grant No.61571296 and N.S.F. of US under Grant No. CNS-1247778, No.
CNS-1619250. Dr. Wen's  work  is partially supported  by  N.S.F. of China under Grant No.61871265.}}

\IEEEtitleabstractindextext{%
\begin{abstract}

This work considers the precoding problem in massive multiuser multiple-input multiple-output (MU-MIMO) systems equipped with low-resolution
digital-to-analog converters (DACs). In previous literature on this topic, it is commonly assumed that the channel state information (CSI)
is perfectly known. However, in practical applications the CSI is inevitably contaminated by noise. In this paper, we propose,
for the first time, an eigen-inference (EI) precoding scheme to improve the error performance of the coarsely quantized
massive MU-MIMO systems under imperfect CSI, which is mathematically modeled by a sum of two rectangular random matrices (RRMs): $\sqrt {1 - \eta } {\bf{H}} + \sqrt \eta  {\bf{E}}$. Instead of performing analysis based on the RRM, using Girko¡¯s Hermitization trick, the proposed method leverages the block random matrix theory by augmenting the RRM into a block symmetric channel matrix (BSCA). Specially, we derive the empirical distribution of the eigenvalues of the BSCA and establish the limiting spectra distribution
connection between the true BSCA and its noisy observation. Then, based on these theoretical results, we propose an EI-based moments matching method for CSI-related noise level ($\eta$) estimation and a rotation invariant estimation method for CSI reconstruction.
Based on the cleaned CSI, the quantized precoding problem is tackled via the Bussgang theorem and
the Lagrangian multiplier method. The prosed methods are lastly verified by numerical simulations and the results demonstrate the effectiveness of the proposed precoder.

\end{abstract}

% Note that keywords are not normally used for peerreview papers.
\begin{IEEEkeywords}
Massive MU-MIMO, Low-Resolution DACs, Eigen-Inference Precoding, Imperfect CSI, Block Random Matrix Theory.
\end{IEEEkeywords}}

% make the title area
\maketitle

\IEEEpeerreviewmaketitle
\IEEEdisplaynontitleabstractindextext

\section{Introduction}
\label{Sec:1}
\IEEEPARstart{R}{ecent} years have witnessed an increasing interest in massive MU-MIMO systems, in which the BS can simultaneously serve a
large number of users equipments (UEs) by utilizing hundreds (even thousands) of transmit antennas \cite{Lu2014An, Larsson2014Massive, Rusek2012Scaling}.
As the number of antennas goes to infinity, the performance of classical precoders, e.g.,
maximum ratio transmission (MRT), zero-forcing (ZF) and water-filling (WF) \cite{Joham2005Linear},
approach to that of optimal nonlinear ones \cite{Jindal2005Dirty}.
On the other hand, gigantic increase in energy consumption, a key challenge to the use of massive MIMO, is brought by scaling up
the number of the antennas. An effective way to decrease the energy cost is to equip the BS with low-resolution analog-to-digital converters (ADCs)
\cite{Zhang2018On} and DACs \cite{Gokceoglu2017Spatio}.

\subsection{Related Works}

The performance of quantized massive MU-MIMO systems has been extensively evaluated in terms of different performance metrics, such as
(coded or uncoded) bit error rate (BER) \cite{Guerreiro2016Use, Zhang2017One, Dong2017Efficient}, coverage probability \cite{Usman2017Joint},
\cite{Yang2017Cell},  and sum rate \cite{Saxena2016On, Mo2017Hybrid, Kong2017Multipair}.  For the simplest case that the massive MU-MIMO are
deployed with 1-bit ADCs
or DACs,  it has been shown in a recent study \cite{Li2017Downlink} that, compared with massive MIMO systems with ideal DACs, the sum rate loss
in 1-bit massive MU-MIMO systems can be compensated by disposing approximately 2.5 times more antennas at the BS. Meanwhile,
 the authors in \cite{Jedda2018On}
presented a quantized constant envelope (QCE) precoder, which can significantly reduce
the power consumption of large-size communication systems.
For frequency-selective MU-MIMO systems,
the achievable performance with 1-bit ADCs or DACs is satisfactory when the number of BS antennas is large
\cite{Jacobsson2017Massive, Studer2016Quantized, Mollen2016Uplink}. The performance gap between 1-bit massive MU-MIMO systems
 and the ideal ones that have access to unquantized data, can be further narrowed by utilizing 3-4 bits ADCs or DACs
\cite{Wen2016Bayes}. %It is shown in quantized constant envelope precoding.
However, coarsely quantized MU-MIMO systems have to
pay a high price for performance loss in terms of BER,
especially when the communication systems are applied with high-order modulations.
%Besides, compared to the ideal linear precoders,
%the BER performance of the quantized linear precoders tend to saturate at a certain signal-to-noise ratio (SNR).

As novel alternatives, quantized nonlinear precoders can significantly outperform linear ones with additional computational complexity.
A novel nonlinear precoder that can support the 1-bit MU-MIMO downlink system with high-order modulations is firstly proposed in
\cite{Jacobsson2016Nonlinear},  which enables 1-bit MU-MIMO system
to work well not only with the QPSK signaling but also with high-order modulations.
Then, the semidefinite relaxation based precoder, which has sound theoretical guarantees and
can achieve a performance close to the optimal nonlinear precoder, has been proposed in \cite{Jacobsson2016Quantized}.  Nevertheless,
its high computational complexity becomes an obstacle to hinder its application in massive MIMO systems.
Several more efficient precoders have been developed recently, such as the
squared-infinity norm Douglas-Rachford splitting based precoder \cite{Jacobsson2016Quantized},
the maximum safety margin precoder \cite{Jedda2016Minimum}, the C1PO and C2PO precoders \cite{Casta20171},
the finite-alphabet precoder \cite{Wang2018Finite}
and the alternating direction method of multipliers based precoder \cite{Chu2018Efficient}. These quantized nonlinear precoders show that, compared to the ideal MU-MIMO (with high-resolution DACs, i.e., 14-bit),  the price paid for the coarsely quantized MU-MIMO is an additional computational complexity.

However, it is worthy of noting that
though the use of low-resolution ADCs and DACs could greatly reduce the power consumption
and maintain performance loss within acceptable margin, the quantized precoders are very sensitive to
the channel estimation error  \cite{Jacobsson2016Quantized}, \cite{Wang2018Finite}, \cite{Chu2018Efficient}.
To the best of our knowledge, the robust precoding scheme for quantized massive MU-MIMO systems in the case of imperfect CSI has not been reported yet.

%In this paper, an EI precoding scheme against imperfect CSI for the coarsely quantized massive MU-MIMO system is developed. Compared to
%the ideal massive MU-MIMO systems that have access to the unquantized signals, it has been shown, using an analytical method,
%in \cite{Guerreiro2018Analytical} that the performance penalty of nonlinear massive MIMO systems with channel estimation errors
%can be bearable if the number of transmit antennas is significantly greater than the number of UEs.
For the downlink of
coarsely quantized massive MU-MIMO systems with imperfect CSI, existing analysis and results on imperfect CSI, e.g.,
\cite{Wong2009Optimal, Ding2010Maximum, Mukherjee2010Robust, Rezk2011On, Lee2012The, Al2016Transmit, miridakis2017on} are not directly applicable
due to the nonlinearity introduced by low-resolution DACs. In this paper, based on the block random matrix theory and
the Bussgang theorem \cite{Julian1952Crosscorrelation},
we develop, for the first time, an EI precoder to address the quantized precoding problem in the case of imperfect CSI.
The main contributions are as follows.

\subsection{Contributions}

Firstly, we provide some theoretical analysis for the channel matrix based on block random matrix theory
\cite{Far2006Spectra, Oraby2007The, Dette2010Random},
which are the theoretical basis of the proposed precoder.
Specially, the rectangular channel matrix is augmented into a block symmetric channel matrix (BSCA) by employing Girko¡¯s Hermitization trick,
 which in principle converts the spectral analysis of rectangular matrices into the spectral analysis of Hermitian matrices.
 Then, we derive the empirical distribution of the eigenvalues of the BSCA based on the block random matrix theory. Meanwhile,
 we establish the limiting spectra distribution connection between the desired BSCA and the noisy BSCA.
 Besides, to facilitate the analysis of the complicated precoding problem, we decompose the basic quantized precoding problem with imperfect
 CSI into three subproblems by using the well-established Bussgang theorem. The decomposed subproblems are then tackled one by one.

Secondly, based on these derived theoretical results, we propose an EI-based moments matching method to
 estimate the unknown CSI-related noise level, and further develop a rotation invariant estimation method
 to reconstruct the CSI from its noisy observation. Based on the reconstructed (refined) CSI,
 the precoding problem is solved via the Lagrangian multiplier method.

Finally, we have evaluated the new precoder in various conditions with imperfect CSI.
The results show that,
with the proposed precoding scheme, significant improvement in robustness against imperfect CSI
can be achieved compared with existing precoders.

\subsection{Paper Outline and Notations}

The remainder of this paper is structured as follows.
Section \ref{IOSM} introduces the system models and outlines the quantized problem for coarsely quantized massive systems
 with channel estimation errors. Section \ref{EIP} presents the basic analysis for the addressed issue and shows, in detail, the proposed
 EI precoder.  In Section \ref{ns}, numerical studies are provided to evaluate the effectiveness
of the proposed EI precoder.
The conclusion of this paper is given in Section \ref{Sec:f}.
For the sake of brevity, the derivations of the technical results are deferred to the Appendices.

\textit{Notations:} Throughout this paper, vectors and matrices are given in
lower and uppercase boldface letters, e.g., $\bf{x}$ and $\bf{X}$, respectively. We use
${\left[ {\bf{X}} \right]_{kl}}$ to denote the element at the $k$th row and $l$th column.
The symbols ${ \mathbb{E} \left[ {\bf{X}} \right] }$, ${ \rm{tr} \left( {\bf{X}} \right) }$, ${ \rm{diag} \left( {\bf{X}} \right) }$,
${\left\| {\bf{X}} \right\|_{\mathop{\rm F}\nolimits} }$, and ${{\bf{X}} ^{\rm H}}$ denote the expectation operator, the trace operator,
the diagonal operator, the Frobenius norm,
the conjugate transpose of ${\bf{X}}$, respectively. $\Re \left( {\bf{x}} \right)$, $\Im \left( {\bf{x}} \right)$,
and ${\left\| {\bf{x}} \right\|_2}$ represent the real part, the imaginary part and
$\ell_2$-norm of vector ${\bf{x}}$.
$\partial f\left(  \cdot  \right)$ stands for the subdifferential of the function $f$.
$\bf{I}$ and $\bf{0}$  are respectively referred to an identity matrix and a zeros matrix with proper size.

\section{SYSTEM MODEL and Problems Formulation}
\label{IOSM}

\subsection{Quantized Massive MU-MIMO System}
We consider a single cell coarsely quantized massive MU-MIMO downlink, operating in a Rayleigh flat-fading environment. In the BS, $N$ antennas simultaneously communicate with $M$ single antenna UEs in the same time-frequency resource.

Let ${\bf{s}} \in {{\mathbb{C}}^{U \times 1}} $ be the constellation points to be sent to UTs. Using the knowledge of CSI, denoted by  ${\bf{\widetilde H}}$, the BS precodes ${\bf{s}}$ into a $N$-dimensional vector ${\bf{x}} = \mathcal{P}\left( {{\bf{\widetilde H}},{\bf{s}}} \right)$, where $\mathcal{P}$ denotes an arbitrary, channel-dependent, mapping between the UT-intended symbols ${\bf{s}}$ and the precoded symbols ${\bf{x}}$. The precoded symbols satisfy the average power constraint \cite{Joham2005Linear},
\begin{equation}
\label{eqi1}
{\mathbb{E}_{\bf{s}}}\left[ {{\bf{x}}^\mathrm{H}}{{\bf{x}}} \right] \le P_{TX}.
\end{equation}

For the quantized MU-MIMO downlink system, each precoded signal component $x_i, \ i = 1, \cdots, N$, is quantized separately into a finite set of prescribed labels by a $B$-bit symmetric uniform quantizer $Q$.
It is assumed that the real and imaginary parts of precoded signals are quantized separately. The resulting quantized signals read
\begin{equation}
\label{eqb2}
{\bf{z}} = Q({\bf{x}}) .
\end{equation}

The input-output relationship of the quantized massive MU-MIMO downlink system can be denoted as
\begin{equation}
\label{eq4}
{\bf{y}} =  {\bf{H}}{\bf{z}} + {\bf{n}},
\end{equation}
where the entries of ${\bf{ H}}$ are complex Gaussian random variables, whose real and imaginary parts are assumed to be independent and identically distributed zero-mean Gaussian random variables with unit variance; ${{\bf{n}}}$ is a complex vector with element $n_i$ being complex addictive Gaussian noise distributed as ${{n_i}} \thicksim \mathcal{CN} \left( {0,{\varepsilon ^2} } \right)$ . The signal-to-noise ratio (SNR) is defined by $SNR  = P_{TX}/{\varepsilon ^2}$.

\begin{figure*}[!ht]
\centering
\subfloat[${\eta = 0, N = 32, M =4}$]{ \label{fig1xxa}
\includegraphics[width=0.33\textwidth]{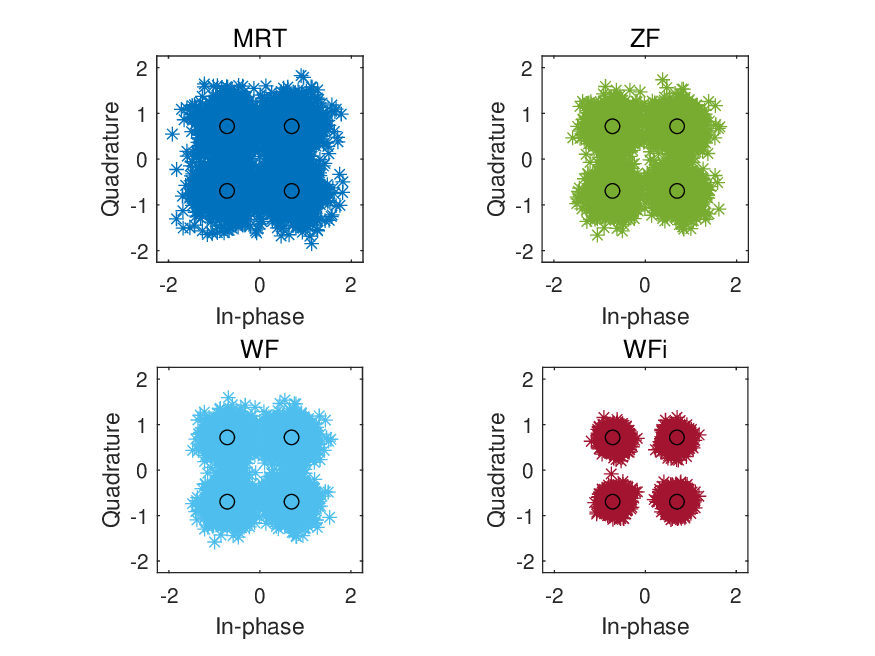}
}
\subfloat[${\eta = 0, N = 128, M =4}$]{ \label{fig1xxb}
\includegraphics[width=0.33\textwidth]{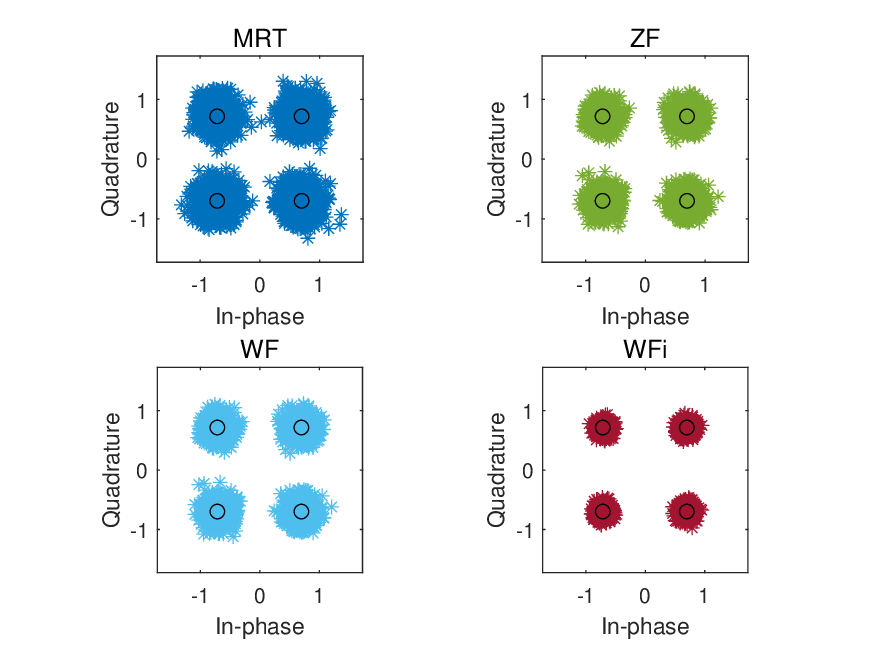}
}
\subfloat[${\eta = 0.36, N = 128, M =4}$]{ \label{fig1xxc}
\includegraphics[width=0.33\textwidth]{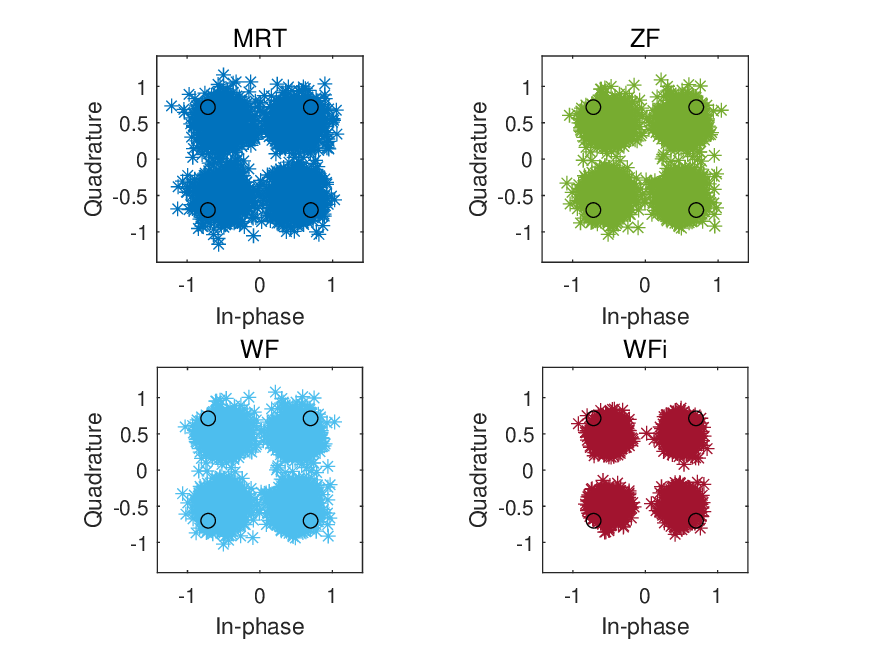}
}
\caption{The estimated outputs for QPSK signaling as a function of CSI-related parameter $\eta$ and system size ($N$ receive antennas and $M$ UEs), at an SNR level of 5dB. }
\label{fig1xx}
\end{figure*}

\subsection{Quantized Precoding Problem Formulation with Imperfect CSI}
\label{icsi}

In previous studies, it is commonly assumed that the CSI is perfectly known, i.e., ${\bf{\widetilde H}} = {\bf{H}}$. However, in practical applications, the CSI is inevitably contaminated by noise. In this paper, we follow \cite{Jacobsson2016Quantized, Al2016Transmit}
for a model of imperfect CSI through a Gauss-Markov uncertainty of the form 
\begin{equation}
\label{eq5}
{\bf{\widetilde H}} = \sqrt {1 - \eta } {\bf{H}} + \sqrt \eta  {\bf{E}},
\end{equation}
where ${{\bf{\widetilde H}}}$ is the imperfect observation of channel available to the BS and $\bf{E}$ is modeled as AWGN. The CSI-related parameter $\eta$ characterizes the partial CSI. Specifically,  $\eta = 0$ means perfect CSI, the values of $0 < \eta  < 1$ 
correspond to partial CSI and $\eta = 1$ accounts for no CSI. In this work, the parameter $\eta$ is restricted to $0 \le \eta  < 1$. 

In the case of imperfect CSI, the quantized precoding problem can be formulated as \cite{Jacobsson2016Quantized}
\begin{equation}
\label{eq6}
\begin{array}{*{20}{c}}
{\mathop {{\rm{minimize}}}\limits_{\beta  \in {\mathbb{R}^ + }} }&{\left\| {{\bf{s}} - \beta {\bf{\widetilde H}}Q\left( {\bf{x}} \right)} \right\|_2^2 + {\beta ^2}M{\sigma ^2}}\\
{{\rm{s}}.{\rm{t}}.}&{E\left\{ {{{\bf{z}}^{\rm{H}}}{\bf{z}}} \right\} \le P_{TX}}
\end{array}.
\end{equation}

We end this part by providing some technical explanations of the model used in \eqref{eq5}. It has shown in numerous studies \cite{Wong2009Optimal, Jacobsson2016Quantized, Al2016Transmit, Wang2018Finite, Caire2007Multiuser, Caire2007Quantized, Guerreiro2018Analytical} that \eqref{eq5} is a widely model to study specific scenarios of the imperfect CSI. The parameter $\beta$ is a function of system parameters depending in different scenarios. For example, with minimum mean-square error channel estimation, $\beta$ represents a function of pilot symbol SNR \cite{Caire2007Multiuser}. For an analog feedback link, $\beta$ is a function of the number of the coefficient of the employed channel and the SNR of the feedback link \cite{Caire2007Quantized}. Recently, $\beta$ is used to concentrate the
differences between the uplink and the downlink SNRs \cite{Guerreiro2018Analytical}. In this paper, we restrict our choice of $\beta$ to a constant value but one can extend the results to being an arbitrary function of the system parameters or investigate the outdated channel state information \cite{Secrecy2018Le}.

\subsection{Why should we need robust precoding for coarsely quantized massive MU-MIMO system?}

For the massive MU-MIMO downlink, there is a growing concern on the excessive energy consumption, which is mainly caused by the data converters at the BS \cite{Lu2014An}.  Equipping the BS with low-resolution DACs (i.e., 1-bit to 4-bit) has been proven to be a potential way to reduce power consumption. Under the assumption of perfect CSI, many studies show that the use of low-resolution DACs could greatly reduce the power
consumption while keeping the performance loss within tolerable levels. However, CSI errors are inevitable in practice.

Fig. \ref{fig1xx} shows the experimental result of the estimated outputs of the MU-MIMO system with 1-bit DACs. Three quantized linear precoding schemes (MRT, ZF, WF,) are employed. Specially, the notation WFi is referred to the case of WF precoding with infinite-resolution DACs. In the case of perfect CSI, it is observed from Fig. \ref{fig1xxa} and Fig. \ref{fig1xxb} that the QPSK constellation becomes more distinguishable with increasing the number of transmit antennas, yielding better BER performance of all precoders. However, Fig. \ref{fig1xxb} and Fig. \ref{fig1xxc} illustrate that the precoding schemes for the MU-MIMO system with low-resolution DACs are more susceptible to CSI errors than that with ideal DACs. In addition, increasing the number of transmit antennas are insufficient in general for securing the BER performance of the coarsely-quantized MU-MIMO systems with imperfect channel knowledge.

The observations described above motivated us to develop a new method for the quantized precoding problem taking into account the CSI errors.

\subsection{Linearized Analysis and Problem Decomposition}

It has been mentioned in Section \ref{Sec:1} and Section \ref{icsi} that there is no available robust algorithm directly applies to solve the quantized precoding problem \eqref{eq6}, which are mainly due to two factors: 1) The nonlinearity of ${\bf x}$ and ${\bf z}$; 2) Imperfect CSI. In what follows, a linearized analysis of \eqref{eqb2} is firstly given to ameliorate the nonlinearity difficulties. Based on the linearized analysis, the original quantized precoding problem is then decomposed into three sub-problems, which are sub-sequently solved.

The proposed algorithm in this paper takes advantage of the Bussgang theorem, and is applicable for any linear precoding scheme, e.g., MRT, ZF, WF, and WFQ \cite{Mezghani2009Transmit}. In what follows, we examine, in detail, the special case of the WF precoder.

Let ${{\bf{\widetilde P}}}$ be a linear precoding matrix. By taking the advantage of the well-known Bussgang theorem \cite{Julian1952Crosscorrelation}, one can obtain
\begin{equation}
\label{eqA1}
{\bf{z}} = Q\left( {\bf{x}} \right) = Q\left( {{\bf{{{\bf{\widetilde P}}}s}}} \right) = {\bf{\widetilde F{{\bf{\widetilde P}}}s}} + {\bf{d}}.
\end{equation}
Plugging \eqref{eqA1} into \eqref{eq6} and applying the Lagrangian multiplier method \cite{Joham2005Linear}, one can get the quantized
WF precoding matrix as:
\begin{align}
\label{eqA2}
{\bf{P}}^{WF} &= \widetilde \beta^{WF} {\bf{\widetilde F}}{\bf{\mathord{\buildrel{\lower3pt\hbox{$\scriptscriptstyle\smile$}}
\over P} }}, \\ {\bf{\mathord{\buildrel{\lower3pt\hbox{$\scriptscriptstyle\smile$}}
\over P} }} &= {{{\bf{\widetilde H}}}^{\rm{H}}}{\left( {{\bf{\widetilde H}}{{{\bf{\widetilde H}}}^{\rm{H}}} + M{\bf \theta }} \right)^{ - 1}}, \notag \\
\widetilde \beta^{WF}  &= \frac{1}{{\sqrt P }}{\mathop{\rm tr}\nolimits} {\left( {{{{\bf{\mathord{\buildrel{\lower3pt\hbox{$\scriptscriptstyle\smile$}}
\over P} }}}^{\rm{H}}}{\bf{\mathord{\buildrel{\lower3pt\hbox{$\scriptscriptstyle\smile$}}
\over P} }}} \right)^{ - 1/2}}, \notag \\
{\bf \theta} &= (\sigma ^2 {\bf{I}} + \sigma_{\bf{d}} ^2),  \notag \\
\sigma _{\bf{d}}^2 &= \left( {1 - {\rm{diag}}\left( {{\bf{\widetilde F}}} \right)} \right)\left( {M{\sigma ^2} + 1} \right). \notag
\end{align}

With results in \eqref{eqA2}, the original precoding problem now can be decomposed into three subproblems: a) Estimation of the CSI-related
parameter $\eta$; b) Reconstruction of CSI  ${\bf{H}}$ from the noisy observation ${\bf{\widetilde H}}$; c) Estimation of the coefficients
matrix $\bf{F}$ under imperfect CSI. The solutions for all the three subproblems are fundamentally based on the block random matrix theory
\cite{Far2006Spectra}, which is presented in details in the following.

%where the noisy WF precoding matrix ${{\bf{\widetilde P}}}$ can be expressed as
%\begin{align}
%{\bf{\widetilde P}} &= \widetilde \beta {\bf{\mathord{\buildrel{\lower3pt\hbox{$\scriptscriptstyle\frown$}}
%\over P} }}, \\ {\bf{\mathord{\buildrel{\lower3pt\hbox{$\scriptscriptstyle\frown$}}
%\over P} }} &= {{{\bf{\widetilde H}}}^{\rm{H}}}{\left( {{\bf{\widetilde H}}{{{\bf{\widetilde H}}}^{\rm{H}}} + M{\sigma ^2}{\bf{I}}} \right)^{ - 1}},
%\notag \\
%\widetilde \beta  &= \frac{1}{{\sqrt P }}{\mathop{\rm tr}\nolimits} {\left( {{{{\bf{\mathord{\buildrel{\lower3pt\hbox{$\scriptscriptstyle\frown$}}
%\over P} }}}^{\rm{H}}}{\bf{\mathord{\buildrel{\lower3pt\hbox{$\scriptscriptstyle\frown$}}
%\over P} }}} \right)^{ - 1/2}}. \notag
%\end{align}
%The problem \eqref{eq6} is challenging since, in practice, we have no prior knowledge of the CSI-related parameter $\eta$. Besides,
%compared with the ideal CSI case, the precoding matrix ${{\bf{\widetilde P}}}$ is corrupted by the environmental noise. In what follows, an
%eigen-inference precoder is proposed to ameliorate this difficulty.

\section{The Proposed Solution: The Eigen-Inference Precoder}
\label{EIP}

We first provide a sketch of the algorithm development
in the following.
%Firstly, by taking advantage of the well-established Bussgang theorem, the original quantized precoding problem \eqref{eq6}
%is decomposed into three subproblems:
%\begin{itemize}
%\item ${\bf{Problem \ I}}$: Estimation of the unknown CSI-related parameter $\eta$;
%\item ${\bf{Problem \ II}}$: Reconstruction of the CSI matrix  ${\bf{H}}$ from the noisy observation ${\bf{\widetilde H}}$;
%\item ${\bf{Problem \ III}}$: Estimation of the coefficients matrix $\bf{F}$ in the case of imperfect CSI.
%\end{itemize}
Firstly, we present some theoretical basics from random matrix theory that will be needed in the ensuing analysis. The main metrics we employ
are introduced and some favorable properties are shown in detail.
Secondly, to estimate the CSI-related parameter $\eta$, an EI-based moments matching
method is then proposed. Furthermore, following the strategies in
\cite{Bun2016Cleaning, Bun2015Rotational}, an EI-based rotation invariant estimator for constructing the
estimation of cleaned CSI ${{\bf{\hat H}}}$ is developed.
Lastly, with the refined CSI, the desired precoding matrix, the cleaned precoding factor, and the
estimate of the coefficient matrix $\bf{\hat F}$ are accordingly obtained.

\subsection{Random Matrix Basics and Main Metrics}
\label{ICK}

We start by presenting some basics from random matrix theory that will be needed in the following algorithm development. For a square random matrix
${\bf{A}} \in {\mathbb{C}}^{M \times M}$, the resolvent of ${\bf{A}}$ is defined as
\begin{equation}
\label{eqa1}
{{\bf{G}}_{\bf{A}}}\left( z \right) = {\left( {z{{\bf{I}}_M} - {\bf{A}}} \right)^{ - 1}}.
\end{equation}
The normalized trace of \eqref{eqa1} gives
\[{\mathfrak{g}}_{\bf{A}}^M\left( z \right) = \frac{1}{M}{\mathop{\rm tr}\nolimits} \left[ {{{\bf{G}}_{\bf{A}}}\left( z \right)} \right].\]
In the limit of large dimension, one has ${\mathfrak{g}}_{\bf{A}}^M\left( z \right)\mathop \rightarrow\limits_{M \to \infty }
{{\mathfrak{g}}_{\bf{A}}}\left( z \right)$, where ${{\mathfrak{g}}_{\bf{A}}}\left( z \right)$ is the $Stieltjes$ transform of ${\bf{A}}$.
The asymptotic empirical distribution of eigenvalues ${F_{\bf{A}}}\left( \lambda  \right)$ (with density ${\rho_{\bf{A}} \left( \lambda  \right)}$) can be described in terms of its $Stieltjes$
transform [22], defined by
\begin{equation}
\label{eqa2}
{{\mathfrak{g}}_{{\bf{A}}}}\left( z \right) = \int {\frac{{d{F_{\bf{A}}}\left( \lambda  \right)}}{{\lambda  - z}}}  = \int {\frac{{\rho_{\bf{A}}
\left( \lambda  \right)}}{{\lambda  - z}}d\lambda } .
\end{equation}
Besides, the $R$ transform, a handy transform which enables the characterization of the limiting eigen-spectra of a sum of free random matrices
 from their individual limiting eigen-spectra, can be defined as
\begin{equation}
\label{eqa3}
{R_{\bf{A}}}\left( z \right) = {\mathfrak{g}}_{\bf{A}}^{ - 1}\left( z \right) - \frac{1}{z}.
\end{equation}
Furthermore, the $R$ transform can be expanded as \cite{Tulino2004Random}:
\begin{equation}
\label{eqaa3}
{R_{\bf{A}}}\left( z \right) = \sum\limits_{l = 1}^\infty  {{\kappa _l}} \left( {\bf{A}} \right){z^{l - 1}},
\end{equation}
where ${\kappa _l}\left( {\bf{A}} \right)$ denotes the so-called $free \ cumulant $ which can be expressed as a function of moments of ${\bf{A}}$.
Specially, given the $m$-th moments of $\bf{A}$ (denoted by $\varphi \left( {{{\bf{A}}^{k}}} \right)$) and using the so-called cumulant formula \cite{Tulino2004Random}, one can have
\begin{equation}
\label{eqaa4}
\varphi \left( {{{\bf{A}}^m}} \right) = \sum\limits_{l = 1}^m {{\kappa _l}\left( {{{\bf{A}}^m}} \right)\sum\limits_{{m_1}, \cdots ,{m_l}}
{\varphi \left( {{{\bf{A}}^{{m_1} - 1}}} \right) \cdots \varphi \left( {{{\bf{A}}^{{m_l} - 1}}} \right)} } ,
\end{equation}
where, for $ {k=1,\cdots,l} $, ${m_k}$ is a non-negative integer and satisfies ${m_1} + ... + {m_l} = m$.
For completeness, the first three cumulants are given by
\begin{equation}
\label{eqaa5}
{\kappa _1} = {\varphi _1}, \ {\kappa _2} = {\varphi _2} - \varphi _1^2, \ {\kappa _3} = {\varphi _3} - 3{\varphi _2}{\varphi _1} + 2\varphi _1^2 ,
\end{equation}
where we denote $\varphi \left( {{{\bf{A}}^{k}}} \right)$ as ${\varphi _k}$ for simplicity.

It is noted that, in our case, ${\bf{H}} \in {\mathbb{C}^{M\times N}}$ is a rectangular random matrix, having complex-valued eigenvalues and
eigenvectors. The well-known Hermitian random matrix theory can not be applied unless a proper strategy is used. In this paper,
our analysis is based on BSCA and the noisy observation of BSCA, of forms
\begin{equation}
\label{eqa4}
{\bf{B}} = \left[ {\begin{array}{*{20}{c}}
{\bf{0}}&{\bf{H}}\\
{{{\bf{H}}^{\mathop{\rm H}\nolimits} }}&{\bf{0}}
\end{array}} \right],
\end{equation}
and
\begin{equation}
\label{eqa5}
{\bf{\widetilde B}} = \left[ {\begin{array}{*{20}{c}}
{\bf{0}}&{{\bf{\widetilde H}}}\\
{{{{\bf{\widetilde H}}}^{\rm{H}}}}&{\bf{0}}
\end{array}} \right] = {\bf{B}} + \alpha \left[ {\begin{array}{*{20}{c}}
{\bf{0}}&{\bf{E}}\\
{{{\bf{E}}^{\rm{H}}}}&{\bf{0}}
\end{array}} \right] = {\bf{B}} + {\bf{\widetilde E}},
\end{equation}
where $\alpha  = \sqrt {\eta /\left( {1 - \eta } \right)} $. The observations in \eqref{eqa4} and \eqref{eqa5} are known as the
Girko's Hermitization trick \cite{Girko1984Circular,Tao2012Topics},
 which has been widely used in the case of theoretical analysis of Hermitian random matrix.
In this paper,
the rectangular random matrix is considered. Thanks to the Girko's Hermitization trick, the theoretical analysis developed in this paper is
 based on the spectral theory of block Hermitian random matrices, instead of the much involved spectral theory of rectangular random matrices.

We, in particular, are aiming to build the connection between the limiting spectra distribution (LSD) of ${\bf{B}}$ and ${\bf{\widetilde B}}$.
Let us define two more auxiliary matrices:
\begin{equation}
\label{eqa6}
{\bf{D}} = {\bf{B}}{{\bf{B}}^{\rm H}} = \left[ {\begin{array}{*{20}{c}}
{{\bf{H}}{{\bf{H}}^{\rm H}}}&0\\
0&{{{\bf{H}}^{\rm H}}{\bf{H}}}
\end{array}} \right],
\end{equation}
and
\begin{equation}
\label{eqa7}
{\bf{\widetilde D}} = {\bf{\widetilde B}}{{{\bf{\widetilde B}}}^{\rm H}} = \left[ {\begin{array}{*{20}{c}}
{{\bf{\widetilde H}}{{\bf{\widetilde H}}^{\rm H}}}&0\\
0&{{\bf{\widetilde H}}{{\bf{\widetilde H}}^{\rm H}}}
\end{array}} \right].
\end{equation}
With notations in \eqref{eqa4}-\eqref{eqa7}, one can find that a direct relationship among eigenvalues of $\bf{B}$ (or ${\bf{\widetilde B}}$),
$\bf{D}$ (or ${\bf{\widetilde D}}$) and ${{\bf{H}}{{\bf{H}}^{\rm H}}}$ (or ${{\bf{\widetilde H}}{{\bf{\widetilde H}}^{\rm H}}}$),
which can be expressed as
\begin{equation}
\label{eqa71}
{\lambda ^{\bf{D}}} = \left\{ {\begin{array}{*{20}{c}}
{{\lambda ^{{\bf{H}}{{\bf{H}}^{\rm H}}}},}&{{\lambda ^{\bf{D}}} \neq 0}\\
{0,}&{{\lambda ^{\bf{D}}} = 0}
\end{array}} \right., \ {\lambda ^{\bf{B}}} =  \pm \sqrt {{\lambda ^{\bf{D}}}} ,
\end{equation}
and
\begin{equation}
\label{eqa72}
{\lambda ^{\bf{\widetilde D}}} = \left\{ {\begin{array}{*{20}{c}}
{{\lambda ^{{\bf{\widetilde H}}{{\bf{\widetilde H}}^{\rm H}}}},}&{{\lambda ^{\bf{\widetilde D}}} \neq 0}\\
{0,}&{{\lambda ^{\bf{\widetilde D}}} = 0}
\end{array}} \right., \ {\lambda ^{\bf{\widetilde B}}} =  \pm \sqrt {{\lambda ^{\bf{\widetilde D}}}}.
\end{equation}
Besides, the limiting spectra connection between the noisy BSCA and the true BSCA can be expressed as in the following results.
\begin{lemma} []
\label{lma3}
For the BSCA ${\bf {B}}$ defined in \eqref{eqa4}, the $Stieltjes$ transform of the LSD of ${\bf{H}}{{{\bf{H}}}^{\rm H}}$ can be represented
by the $Stieltjes$ transform of the LSD of ${\bf {B}}$ as:
\begin{equation}
\label{eqthma3}
{{\mathfrak{g}}_{{\bf{H}}{{\bf{H}}^H}}}\left( z \right) = \frac{{q + 1}}{{2q}}{{\mathfrak{g}}_{\bf{D}}}\left( z \right) -
\frac{{q - 1}}{{2q}}\frac{1}{z}
\end{equation}
\end{lemma}
Lemma \ref{lma3} bridges up the LSD connections between the interested block random matrix ${\bf{D}}$ and ${{\bf{H}}{{\bf{H}}^{\rm H}}}$,
whose density is subject to the well-known Marchenko-Pastur law.
\begin{lemma} []
\label{lma1}
Let $q = M/N < 1$. For the BSCA ${\bf{B}}$ and the sample covariance of the BSCA ${\bf{D}}$, the $Stieltjes$
 transforms of the LSDs of ${\bf{B}}$ and ${\bf{D}}$ satisfy
\begin{equation}
\label{eqthla1}
{{\mathfrak{g}}_{\bf{B}}}\left( z \right) = z{{\mathfrak{g}}_{\bf{D}}}\left( {{z^2}} \right).
\end{equation}
\end{lemma}

\begin{lemma} []
\label{lma2}
Let the random matrices ${\bf{D}}$ and ${\bf{B}}$ be defined as in Lemma \ref{lma1}. The $S$ transforms of the LSDs of ${\bf{B}}$ and ${\bf{D}}$ meet
\begin{equation}
\label{eqthma2}
{\left[ {{S_{\bf{B}}}\left( z \right)} \right]^2} = \frac{{1 + z}}{z}{S_{\bf{D}}}\left( z \right).
\end{equation}
\end{lemma}
Lemma \ref{lma1} and Lemma \ref{lma2} state direct LSD linkage between ${\bf{B}}$ and ${\bf{D}}$.
The proof of the above lemmas are respectively deferred to Appendix \ref{APX1a}, Appendix \ref{APX1b}, and Appendix \ref{APX1c}.
 With the technical results shown in Lemma \ref{lma3}, Lemma \ref{lma1} and Lemma \ref{lma2}, the statistical properties
 (i.e., the asymptotic empirical distribution of eigenvalues) of ${\bf{B}}$ can be obtained by the well-established
 Hermitian random matrix theory. Besides, combined with the addition law \cite{Tulino2004Random, Couillet2012Random}:
\begin{equation}
\label{eqa73}
{R_{{\bf{X}} + {\bf{Y}}}}\left( z \right) = {R_{\bf{X}}}\left( z \right) + {R_{\bf{Y}}}\left( z \right),
\end{equation}
one can reveal the limiting spectra connection between the noisy BSCA and the true BSCA.

All technical results shown above act as fundamental roles which enable us to tackle the quantized precoding problem introduced in Section \ref{icsi}.
The detailed analysis is given in the following.

\subsection{Determination of the Unknown Parameter $\eta$: An EI-based Moments Matching Algorithm}

In this part, we propose a moments matching algorithm to determinate the unknown parameter $\eta$ in \eqref{eq5}. Our analysis is based
on the free probability which is a powerful tool for analyzing the eigen-spectra of large random matrices. The proposed moments matching algorithm
can provide a robust and flexible estimation of $\eta$ without specifying an explicit relationship between the number of UEs and
the number of antennas.

The estimation of $\eta$ is enabled by random matrix theory which offers a collection of useful results on the asymptotic behavior of the
BSCA $\bf{B}$ and its noisy observation ${\bf{\widetilde B}}$, which are shown in the following theoretical results.

\begin{figure}[!ht]
\centering
\includegraphics[width=0.48\textwidth]{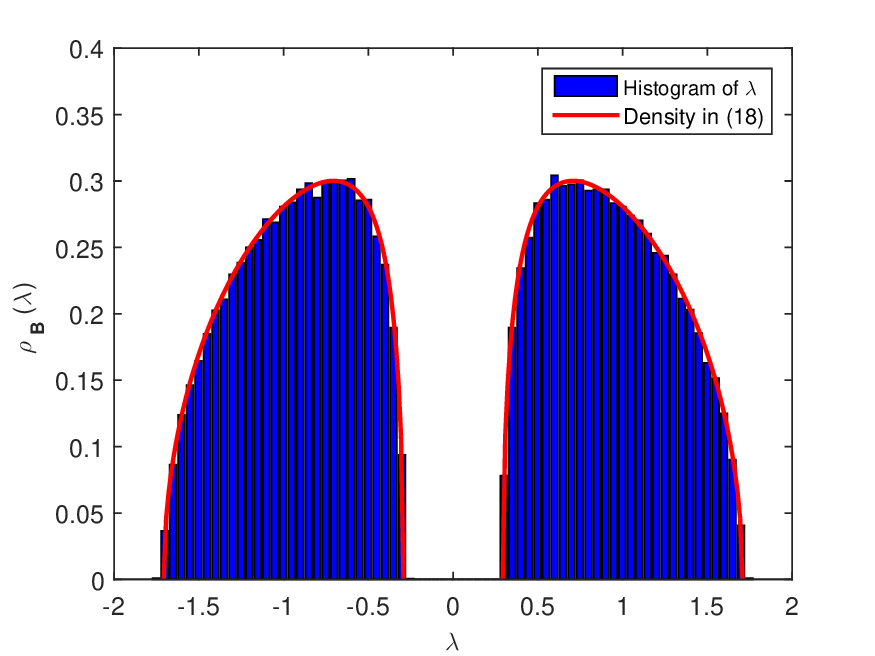}
\caption{The empirical distribution of the nonzero eigenvalues of BSCA: $M=128$, $N=256$, $q=0.5$.}
\label{fig1}
\end{figure}

\begin{myTheo} []
\label{thmb1}
For the BSCA $\bf{B}$ defined in \eqref{eqa4}, the empirical distribution of the eigenvalues of $\bf{B}$ converges almost surely,
as $M,N \rightarrow \infty $ with $M/N \rightarrow q$, to a nonrandom distribution whose density function is
\begin{equation}
\label{eqthm1}
{\rho _{\bf{B}}}\left( x \right) = \frac{{\sqrt {\left( {{b^2} - {x^2}} \right)\left( {{x^2} - {a^2}} \right)} }}
{{\left( {q + 1} \right)\pi \left| x \right|}} + \frac{{1 - q}}{{1 + q}}\delta \left( x \right),
\end{equation}
where $a \le \left| x \right| \le b$ and \[a = 1 - \sqrt q ,b = 1 + \sqrt q .\]
\end{myTheo}
The proof of Theorem \ref{thmb1} is deferred to Appendix \ref{APX2}. As an illustration of the result obtained in Theorem \ref{thmb1},
Fig. \ref{fig1} shows the empirical distribution of the nonzero eigenvalues of the BSCA. it can been seen that the
empirical density function \eqref{eqthm1} agrees well with the simulation result.

Combined with the favorable results given in Theorem \ref{thmb1}, Lemma \ref{lma1}, Lemma \ref{lma2} and Lemma \ref{lma3},
one can obtain the following result for the sample covariance of the noisy channel matrix.

\begin{myTheo}
\label{thmb2}
For the noisy channel matrix ${\bf {\widetilde H}}\in {\mathbb{C}}^{M \times N}$,
the empirical distribution of eigenvalues of its sample covariance
matrix ${{\bf{\widetilde H}}{{\bf{\widetilde H}}^{\rm H}}}$ converges almost surely to a limit distribution whose $Stieltjes$ transform
, denoted by $G={{\mathfrak{g}}_{{\bf{\widetilde H}}{{\bf{\widetilde H}}^{\rm H}}}}\left( z \right)$, satisfies
\label{coly}
\begin{equation}
\label{eqthm3}
{\rm{ - 2}}Gz + {G^2}\left( {1 + {k^2}} \right)qz + h\left( G \right) = 0,
\end{equation}
as $M,N \to \infty$ with the ratio $q=M/N<1$ fixed, and
\begin{align}
h\left( G \right) &= \sqrt {1 + {G^2}z\left( {4 + q\left( { - 2 + {G^2}qz} \right)} \right)}   \notag \\
 &+ \sqrt {1 + {G^2}{\alpha^2}z\left( {4 + q\left( { - 2 + {G^2}z{\alpha^2}q} \right)} \right)}.  \notag
\end{align}
\end{myTheo}
The proof of Theorem \ref{thmb2}  is deferred to Appendix \ref{APX3}. Combined the theoretical result in Theorem \ref{coly}
with the relation in \eqref{eqa3}, we can obtain the $R$ transform of ${{\bf{\widetilde H}}{{\bf{\widetilde H}}^{\rm H}}}$.
Accordingly, we can infer the parameters of the underlying noisy channel matrix  ${\bf{\widetilde H}}$  from a realization of
its sample covariance matrix.  Taking Taylor's expansion for the obtained $R$ transform, the first three
moments\footnote{Here we only employ the first three moments by considering the trade-off between the computational complexity and
estimation accuracy. See more details in Section \ref{upeta}.}
of the sample covariance matrix can be analytically parameterized by the unknown parameters
$\eta$ of form:
\begin{align}
\varphi \left( {{\bf{\widetilde H}}{{\bf{\widetilde H}}^{\rm H}}} \right) &= \frac{1}{{1 - \eta }},\\
\varphi \left[ {{{\left( {{{\bf{\widetilde H}}{{\bf{\widetilde H}}^{\rm H}}}} \right)}^2}} \right] &= \frac{{2\left( {  1 - \eta} \right)
{\eta}\left( { 1 - q} \right) + q}}{{{{\left( {  1 - \eta} \right)}^2}}}, \\
\varphi \left[ {{{\left( {{{\bf{\widetilde H}}{{\bf{\widetilde H}}^{\rm H}}}} \right)}^3}} \right] &= \frac{{q\left( {3\left(
{1 - \eta} \right){\eta}\left( {1 - q} \right) + q} \right)}}{{{{\left( {1 - \eta} \right)}^3}}}.
\end{align}

Given an observation ${\bf{\widetilde H}}$, we can compute estimates of the first three moments of its sample covariance matrix as
\cite{Tulino2004Random}
\begin{equation}
\label{e99}
\hat \varphi \left[ {{{\left( {{{\bf{\widetilde H}}{{\bf{\widetilde H}}^{\rm H}}}} \right)}^k}} \right] = \frac{1}{M}{\mathop{\rm tr}\nolimits}
\left[ {{{\left( {{{\bf{\widetilde H}}{{\bf{\widetilde H}}^{\rm H}}}} \right)}^k}} \right],
\end{equation}
for $k = 1,2,3.$ Since $q=N/M$ is already known, we can estimate the CSI-related parameter ${\eta }$ by EI-based $moments \ matching$,
 in particular, by solving the non-linear system of equations:
\begin{equation}
\label{e100}
\hat \alpha  = \mathop {{\mathop{\rm argmin}\nolimits} }\limits_{\alpha  > 0} {\left\| {\sum\limits_{k = 1}^3 {\hat \varphi \left[
{{{\left( {{{\bf{\widetilde H}}{{\bf{\widetilde H}}^{\rm H}}}} \right)}^k}} \right] - \varphi \left[ {{{\left(
{{{\bf{\widetilde H}}{{\bf{\widetilde H}}^{\rm H}}}} \right)}^k}} \right]} } \right\|^2}.
\end{equation}
Though the EI-based moments matching algorithm described above is theoretically exact when $M,N \rightarrow \infty$,
we verify through various simulations
that the proposed method is sufficiently accurate for even small values of $N$ and $M$ (i.e., $M = 8$, $N = 32$).
See experimental results in Section \ref{ns} for more details.

\subsection{CSI Cleaning Based on Random Matrix Theory: EI-based Rotation Invariant Estimation}
\label{Section3C}

We now attempt to construct an estimator ${{\bf{\hat H}}}$ of the true channel matrix ${\bf{H}}$ that relies on the imperfect observation
${{\bf{\widetilde H}}}$. We will focus on the case that ${\bf{H}}$ is an $M \times N$ random matrix that ${M,N} \to \infty$ with the ratio $q=M/N<1$ fixed. Our analysis is based on the rotation invariant estimation theory \cite{Takemura1984An,Bun2016Cleaning, Bun2015Rotational} and by leveraging spectral properties of the BSCA (shown in Section \ref{ICK}). 

Let ${\lambda _1} \ge  \cdots  \ge {\lambda _{M + N}}$ and ${{\bf{u}}_1}, \cdots ,{{\bf{u}}_{M + N}}$ be the eigenvalues and eigenvectors of a true matrix (denoted by $\bf{A}$), respectively. For the noisy matrix (denoted by ${\bf{\widetilde A}}$), its eigenvalues and eigenvectors are respectively denoted by ${\omega _1} \ge  \cdots  \ge {\omega _{M + N}}$ and ${{\bf{v}}_1}, \cdots , {{\bf{v}}_{M + N}}$.

Based on the minimum mean-square error (MMSE) criterion \cite{kay1993fundamentals}, one can obtain the optimal estimate of $\bf{A}$ by solving
\begin{equation}
\label{eq3c3}
{\bf{\hat A}} = \mathop {\arg {\rm{minimize}}}\limits_{\Omega ({\bf{\tilde A}})} \left\| {\Omega ({\bf{\tilde A}}) - {\bf{A}}} \right\|_{\rm{F}}^2.
\end{equation}
In this paper, we use ${\Omega ({\bf{\tilde A}})}$ to denote the set of all possible rotation invariant estimators. In statistics (see the tutorial work of \cite{Bun2016Cleaning} for more details.), any rotation invariant estimator $\Omega \left( {\bf{A}} \right)$ enjoys the same eigenvectors as the perturbed matrix ${\bf{\widetilde A}}$ . In other words, we can have
\begin{equation}
\label{eq3c4}
\Omega \left( {\bf{A}} \right) = \sum\limits_{k = 1}^{M + N} {{\xi _k}{{\bf{v}}_k}{\bf{v}}_k^{\mathop{\rm H}\nolimits} }, 
\end{equation}
where ${\xi _1}, \cdots, {\xi _{M+N}}$ are quantities to be determined. Using the fact that $\left\| {\bf{W}} \right\|_{\mathop{\rm F}\nolimits} ^2 = {\mathop{\rm tr}\nolimits} \left( {{\bf{W}}{{\bf{W}}^{\rm{H}}}} \right)$, one can have
\begin{equation}
\label{equtk}
\left\| {\Omega ({\bf{\tilde A}}) - {\bf{A}}} \right\|_{\rm{F}}^2 = {\bf{tr}}{\left( {\sum\limits_{k = 1}^{M + N} {{\xi _k}{{\bf{v}}_k}{\bf{v}}_k^{\rm{H}} - \sum\limits_{k = 1}^{M + N} {{\lambda _k}{{\bf{u}}_k}{\bf{u}}_k^{\rm{H}}} } } \right)^2}.
\end{equation}
Since ${\xi _k}, k=1,\cdots,M+N$ are quadratic in \eqref{equtk},  substituting \eqref{eq3c4} into \eqref{eq3c3} gives the optimal solution of \eqref{eq3c3}:
\begin{equation}
\label{eq3c5}
{\bf{\hat A}} = \sum\limits_{k = 1}^{M + N} {{\hat \xi _k}{{\bf{v}}_k}{\bf{v}}_k^{\mathop{\rm H}\nolimits} },
\end{equation}
and
\begin{equation}
\label{eq3c6}
{{\hat \xi }_k} = \sum\limits_{j = 1}^{M + N} {{\lambda _j}{{\left( {{\bf{u}}_j^{\rm{H}}{{\bf{v}}_k}} \right)}^2}}.
\end{equation}
In the limit of large dimensions \cite{schwarz1978estimating}, ${{\hat \xi }_k}$ can be approximately calculated by its expectation value 
\[{{\hat \xi }_k} = \sum\limits_{j = 1}^{M + N} {{\lambda _j}\mathbb{E}\left[ {{{\left( {{\bf{u}}_j^{\rm{H}}{{\bf{v}}_k}} \right)}^2}} \right]}. \]

%where
%\begin{equation}
%\label{eq3c7}
%O\left( {{\omega _k},{\lambda _j}} \right) = (M+N) \mathbb{E}\left[ {{\bf{u}}_k^{\mathop{\rm H}\nolimits} {{\bf{v}}_j}} \right].
%\end{equation}
%The expression $O\left( {{\omega _k},{\lambda _j}} \right)$ in \eqref{eq3c7} is so-called $miracle$ overlap which creates
%deep connection between the true eigenvalues ${\lambda _j}$ and perturbed eigenvalues ${\omega _k}$. By the $miracle$ overlap,
%in large dimensional regime, one can estimate ${\lambda _j}$ only from the measurements of LSD of ${\bf{\widetilde A}}$.
In particular, let ${\bf{\widetilde A}} = {{\bf{\widetilde H}}{{\bf{\widetilde H}}^{\rm H}}}/\left( {1 - \eta } \right)$, based on rigorous  mathematical derivation, one can obtain the estimates of
the true eigenvalues of ${\bf A} = {{{\bf{H}}^{\rm H}}{\bf{H}}}$ by \cite{Bun2015Rotational}
\begin{equation}
\label{eq3c8}
{{\hat \lambda }_k} = {\phi _1}\left( {{\omega _k}} \right){\phi _2}\left( {{\omega _k}} \right) + {\phi _3}\left( {{\omega _k}} \right),
\end{equation}
where
\begin{align}
{\phi _1}\left( {{\omega _k}} \right) &= 1 - q{\alpha ^2}{h_k}, \alpha  = \sqrt {\eta /\left( {1 - \eta } \right)} ,\notag\\
{\phi _2}\left( {{\omega _k}} \right) &= {{\omega _k} - {\alpha ^2}\left( {1 - q} \right) - 2q{\alpha ^2}{h_k}} , \notag\\
{\phi _3}\left( {{\omega _k}} \right) &= {h_k}\left( {{\omega _k} - {\alpha ^2}\left( {1 - q} \right)
+ q{\alpha ^2}{\omega _k}\left( {{\rho _k} - {h_k}} \right)} \right), \notag \\
{\rho _k} &= \mathop {\lim }\limits_{{\epsilon _k} \to {0^ + }} {\mathop{\Im}\nolimits}
\left[ {{\mathfrak{g}_{\bf{A}}}\left( {{\omega _k} + {\mathop{\rm i}\nolimits} {\epsilon _k}} \right)} \right] , \notag \\
{h_k} &= \mathop {\lim }\limits_{{\epsilon _k} \to {0^ + }} {\mathop{\Re}\nolimits} \left[ {{\mathfrak{g}_{\bf{A}}}
\left( {{\omega _k} + {\mathop{\rm i}\nolimits} {\epsilon _k}} \right)} \right]. \notag
\end{align}

With \eqref{eq3c8} and the direct relation shown in \eqref{eqa71} and \eqref{eqa72}, one can get the estimates of the eigenvalues
of $\bf{D}$ and $\bf{B}$. With \eqref{eq3c5}, we can obtain the estimates of $\bf{D}$ and $\bf{B}$ (denoted by $\bf{\hat B}$). Finally,  the estimate of the true channel matrix is the upper nonzero block of $\bf{\hat B}$.

The results in \eqref{eq3c8} have been proven by means of the Replica method \cite{Bun2016Cleaning, Bun2015Rotational} in the case of
the Hermitian random matrix. The analysis performed above is based on the theoretical work of \cite{Bun2016Cleaning, Bun2015Rotational}.
Our contributions are:1) We develop an EI-based
moments matching method to estimate the CSI-related parameter $\eta$, which is actually unknown in practice. 2) We
propose a theoretically simple approach to derive the estimator of the $Stieltjes$ transform of ${\bf{\widetilde A}}$ (denoted by ${\mathfrak{g}_{\bf{A}}}(z)$) in Theorem \ref{thmb2}, which is useful in \eqref{eq3c8}.

\subsection{Determination of the Coefficient Matrix $\bf{F}$}

With cleaned CSI, we are now ready to determine $\bf{F}$. It has been widely assumed that the elements of the uniform quantization
errors $\bf{d}$ converge in distribution to a zero-mean Gaussian random variable, whose variance can be characterized in closed form.
Indeed, such an assumption is essentially valid when the errors from individual channels are asymptotically pairwise independent,
each uniformly distributed within the quantization thresholds \cite{Jimenez2015White}. Denote the estimated precoding matrix by ${ \bf{\hat P}}$.
With Gaussian assumption of ${\bf{d}}$ and based the Bussgang theorem, one can estimate $\bf{F}$ by minimzing the square error
\begin{equation}
\label{eqE1}
\mathbb{E}\left\| {\bf{d}} \right\|_2^2 = \mathbb{E}\left\| {{\bf{x}} - {\bf{F z}}} \right\|_2^2,
\end{equation}
where ${\bf z} = {\bf {\hat P}s}$. Then
\begin{equation}
\label{eqE2}
\begin{array}{l}
\begin{aligned}
{\bf{\hat F}} &= \mathop {\arg {\rm minimize} }\limits_{\bf{F}} \mathbb{E}\left\| {{\bf{x}} - {\bf{Fz}}} \right\|_2^2\\
 &= \mathop {\arg {\rm minimize} }\limits_{\bf{F}} {\rm{tr}}\left( {{{\bf{R}}_{{\bf{xx}}}} - {{\bf{R}}_{{\bf{xz}}}}{{\bf{F}}^{\rm{H}}} -
 {\bf{F}}{{\bf{R}}_{{\bf{zx}}}} + {\bf{F}}{{\bf{R}}_{{\bf{zz}}}}{{\bf{F}}^{\rm{H}}}} \right)\\
 &= \mathop {\arg {\rm minimize} }\limits_{\bf{F}} f\left( {\bf{F}} \right),
\end{aligned}
\end{array}
\end{equation}
can be obtained. A solution of \eqref{eqE2} has to satisfy
\begin{equation}
\label{eqE3}
\frac{{\partial f\left( {\bf{F}} \right)}}{{\partial {\bf{F}}}} =  - 2{{\bf{R}}_{{\bf{xz}}}} + 2{\bf{F}}{{\bf{R}}_{{\bf{zz}}}},
\end{equation}
which yields
\begin{equation}
\label{eqE4}
{\bf{\hat F}} = {{\bf{R}}_{{\bf{xz}}}}{\bf{R}}_{{\bf{zz}}}^{ - 1},
\end{equation}
where
\begin{equation}
\label{eqE5}
{{\bf{R}}_{{\bf{zz}}}} = {\mathbb{E}_{\bf{s}}}\left[ {{\bf{z}}{{\bf{z}}^{\rm{H}}}} \right] = {\bf{\hat P}}
{\mathbb{E}_{\bf{s}}}\left[ {{\bf{s}}{{\bf{s}}^{\rm{H}}}} \right]{{\bf{\hat P}}^{\rm{H}}} = {\bf{\hat P}}{{\bf{\hat P}}^{\rm{H}}},
\end{equation}
and
\begin{equation}
\label{eqE6}
{{\bf{R}}_{{\bf{xz}}}} = {\mathbb{E}_{\bf{s}}}\left[ {{\bf{x}}{{\bf{z}}^{\rm{H}}}} \right] = {\mathbb{E}_{\bf{s}}}\left[
{Q\left( {\bf{z}} \right){{\bf{z}}^{\rm{H}}}} \right].
\end{equation}
It is assumed that $\bf{F}$ is a diagonal matrix\footnote{This assumption, despite its simplicity, is a common assumption
and proven to be accurate in previous works \cite{Jacobsson2016Quantized, Casta20171, Li2017Downlink, Saxena2017Analysis} and
verified in our experimental results in Sec. \ref{ns}.}, then we can obtain its diagonal elements
\begin{equation}
\label{eqE7}
{\left[ {{\bf{\hat F}}} \right]_{m,m}} = \frac{{{E_x}\left[ {Q\left( x \right){x^ * }} \right]}}{{\sigma _m^2}},
\end{equation}
where $m = 1,2, \cdots M$ and $\sigma _m^2 = {\left[ {{{\bf{R}}_{{\bf{zz}}}}} \right]_{m,m}}$.

Let \[{l_b} = \Delta \left( {b - \frac{{{2^B} - 1}}{2}} \right), \  b = 1, \cdots ,{2^B - 1}\ ,\] and \[{\tau _b} =
\Delta \left( {b - {2^{B - 1}}} \right), \ b = 2, \cdots ,{2^B} \ ,\] respectively be the quantization labels and
quantization thresholds of a $B$ bit uniform quantizer, whose output can be expressed as
\begin{equation}
\label{eqE8}
Q\left( z \right) = \frac{\Delta }{2}\left( {1 - {2^B}} \right) + \Delta \sum\limits_{l = 1}^{{2^B} - 1}
{{\delta_{\left[ {\Delta \left( {l - {2^{B - 1}}} \right),\left. \infty  \right)} \right.}}\left( z \right)},
\end{equation}
where $\delta_{\Phi} \left( {x} \right)$ is an indicator function defined as
\[{\delta _\Theta }\left( x \right) = \left\{ {\begin{array}{*{20}{c}}
1&{{\rm{if } \ }x \in \Theta }\\
0&{{\rm{if } \ }x \notin \Theta }
\end{array}} \right. .\]
Substituting \eqref{eqE8} into \eqref{eqE7} gives
\begin{equation}
\label{eqE9}
\begin{array}{c}
\begin{aligned}
{\left[ {{\bf{\hat F}}} \right]_{m,m}} &= \frac{\Delta }{{\sigma _u^2/2}}\sum\limits_{l = 1}^{{2^B} - 1}
{\int_{l - {2^{B - 1}}}^\infty  {\frac{x}{{\sqrt {\pi \sigma _u^2} }}\exp \left( { - \frac{{{x^2}}}{{\sigma _u^2}}} \right)dx} } \\
 &= \frac{\Delta }{{\sqrt {\pi \sigma _u^2} }}\sum\limits_{l = 1}^{{2^B} - 1} {\exp \left(
  { - \frac{{{\Delta ^2}}}{{\sigma _u^2}}{{\left( {l - {2^{B - 1}}} \right)}^2}} \right)}
\end{aligned}
\end{array}.
\end{equation}

So far, the quantized precoding problem has been tackled. The proposed EI precoder is summarized in Algorithm \ref{algorithm}.

\begin{algorithm}[!ht]
    \caption{The Proposed EI Precoding Algorithm}
    \begin{algorithmic}[1]
    \label{algorithm}
        \STATE $\bf{Input:}$ Imperfect SCM of CSI: ${\bf{\widetilde H}}$.
        \STATE Construct noisy BSCA ${\bf{\widetilde B}}$ using \eqref{eqa5}.
        \STATE Calculate the estimates of the first three moments of
        ${{\bf{\widetilde H}}{{\bf{\widetilde H}}^{\rm H}}}$
         using \eqref{e99} and compute the related moments using \eqref{eqaa5}.
        \STATE Substituting the obtained moments
        $\hat \varphi \left[ {{{\left( {{{\bf{\widetilde H}}{{\bf{\widetilde H}}^{\rm H}}}} \right)}^k}} \right]$ into \eqref{e100}
         to estimate the CSI-related parameter ${\hat \eta }$.%
        \STATE Performing eigenvalue decomposition of ${\bf{\widetilde B}}$ yields
        \[ \left[ {{\bf{U}},{\bf{S}}} \right] = {\mathop{\rm eig}\nolimits} \left( {\bf{\widetilde B}} \right).\]
        \STATE Combined with the CSI-related parameter ${\hat \eta }$, calculating the estimates of eigenvalues of ${\bf{B}}$
        using  \eqref{eqa72} and \eqref{eq3c8} gives $\hat {\bf{S}}$.
        \STATE Reconstructing ${\bf{B}}$ as
        \[{\bf{\hat B}} = {{{\bf{U}} \hat {\bf{S}}}}{{\bf{U}}^ {\rm{H}}}.\]
        The estimated CSI, denoted by ${\bf{\hat H}}$, is the upper triangular part of ${\bf{\hat B}}$ and  ${\bf{\hat {H}}}
        = {{{\bf{\hat B}}}_{1:M,M + 1:M + N}}$.
        \STATE Plugging the estimated CSI ${\bf{\hat H}}$ into \eqref{eqA2} yields the estimates of the precoding matrix and
        precoding factor, which are denoted by $\bf{\hat P}$ and $\bf{\hat \beta}$, respectively.
        \STATE Calculating the estimate of the coefficients matrix $\bf{F}$ using \eqref{eqE9} to obtain $\bf{\hat F}$.
        \STATE $\bf{Output:}$ $\bf{\hat P}$, $\bf{\hat F}$, and $\bf{\hat \beta}$.
    \end{algorithmic}
\end{algorithm}

\subsection{Complexity Analysis}

We now analyze the computational complexity of the proposed algorithms (Algorithm \ref{algorithm}). In step 3 of Algorithm \ref{algorithm}, computing the first three moments has a complexity of $O\left(k_1MN^2\right)$, which involves the cost of $k_1$ matrix multiplications. Step 4 of Algorithm \ref{algorithm} is the linear square problem with a closed form solution in the form of pseudoinverse, which yields complexity of $O\left(K^w\right)$ ($K$ is the number of the unknown parameter and $w$ satisfies $w < 2.37$ \cite{Stothers2010On}).  The complexity of the eigenvalue decomposition in Step 5 is $O\left(M^3\right)$. In addition, the computation of Step 6-9 requires $O\left(k_2MN^2+k_3M^3\right)$ multiplications, which involve the cost of computing matrix-vector multiplication/addition and the Cholesky factorization (in the matrix inverse). The overall complexity required by the proposed precoder is $O\left((k_1+k_2)MN^2+(k_3+1)M^3+K^w\right)$.  Compared to the classical linear quantized precoders \cite{Jacobsson2017Massive, Jedda2016Minimum}, the proposed eigen-inference precoder only incurs slightly more computational complexity (several matrix multiplications and one eigenvalue decomposition).

\section{Numerical Studies}
\label{ns}

This section evaluates the performance of the new precoder via numerical simulations, which include the evaluation of the parameter $\eta$ estimation, CSI reconstruction, and BER performance. For BER performance comparison, we compare the proposed precoder with five representative linear quantized precoders: MRT, WF, WFQ, ZF, and QCE. Each provided result is an average over 1000 independent runs.

\subsection{The Estimation Accuracy of the Unknown Parameter $\eta$}
\label{upeta}

In the first experiment, we investigate the estimation accuracy of the unknown CSI-related parameter $\eta$ in three different settings,
 in terms of the empirical cumulative distribution function (CDF) \cite{Kun2018Fun} of the estimation error \footnote{Due to the nonlinearity of low-resolution DACs, we cannot obtain an exact expression of theoretical CDF of the estimation error. As a result, we investigate the empirical CDF of the estimation error.}.
The results shown in the following are the CDFs of the absolute estimation error, which is denoted by
\[\Delta \eta {\rm{ = }}\left| {\eta {\rm{ - }}\hat \eta }  \right|,\]
where $\hat \eta$ is the estimate of $\eta$.
%The results showing in this subsection are the obtained by 1000 independent runs.

\subsubsection{The Effect of the Order of Moments}

\begin{figure}[!ht]
\centering
\includegraphics[width=0.48\textwidth]{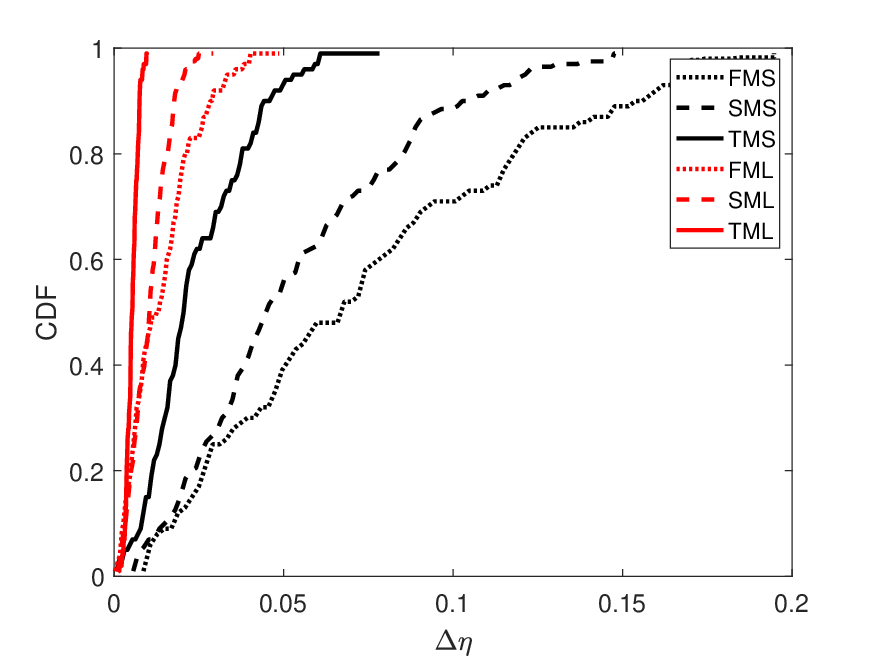}
\caption{The CDFs of the absolute estimation error in the case of different order of the moments.}
\label{fig5}
\end{figure}

The result shown in Fig. \ref{fig5} is obtained by numerically solving \eqref{e100}. The selected root of \eqref{e100} satisfies $0<\eta<1$.
The true value of $\eta$ is fixed at 0.5.
Fig. \ref{fig5} shows the CDFs of the absolute estimation error $\Delta \eta$ in the case of different order of the moments estimates.
In the case
of a small-size MU-MIMO system (with $M$=32, $N$ = $8$), FMS, SMS, and TMS are respectively referred to first-order, second-order,
and third-order of moments
estimates ($k=1 \ or \ 2 \ or \ 3$ in \eqref{e100}). Similarly, FML, SML, and TML mean first-order, second-order, and third-order of
moments estimates in
a large-scale  MU-MIMO system ($M$=256, $N$ = $30$), respectively. From Fig. \ref{fig5}, we can see that the estimation error
decreases as the order
of the moments increases. For the large-size MU-MIMO system, the maximum of absolute estimation error is less than 0.05, which is only $10\%$
of $\eta$, indicating the robustness of the proposed method. Besides, compared to the small-size MIMO system,
 better estimation performance can be obtained in MU-MIMO systems with large-size transmit antennas.
However, it is noted that employing higher order of moments pays the price of higher computational complexity.
In what follows, we respectively
use first-order and third-order moments estimates to determine the unknown parameter $\eta$ in small-size ($M\leq32$)
 and large-size ($M\geq128$) systems.

\subsubsection{The Effect of the System Size}
\label{ess}

\begin{figure}[!ht]
\centering
\includegraphics[width=0.48\textwidth]{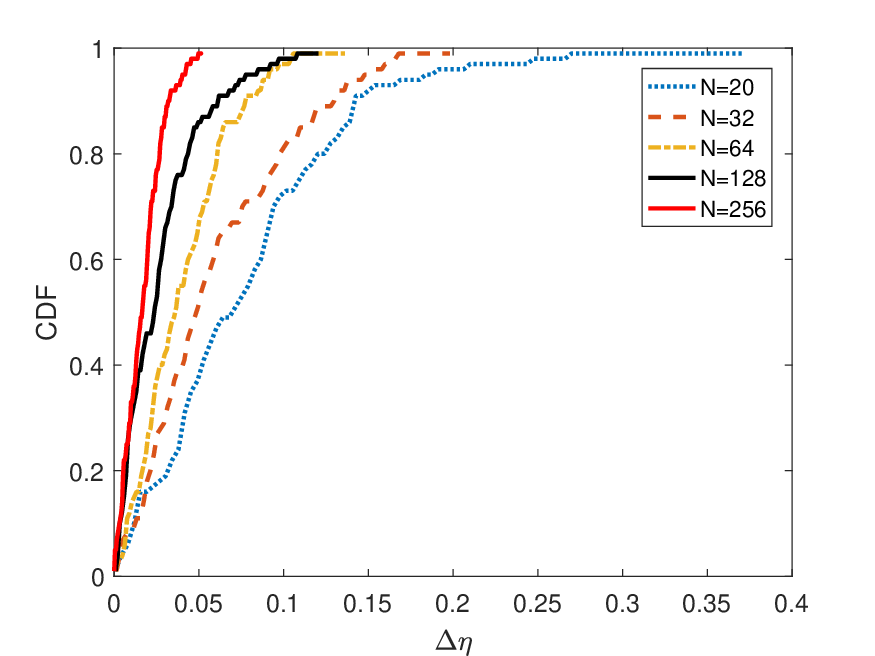}
\caption{The CDFs of the absolute estimation error in the case of different number of transmit antennas.}
\label{fig4a}
\end{figure}

\begin{figure}[!ht]
\centering
\includegraphics[width=0.48\textwidth]{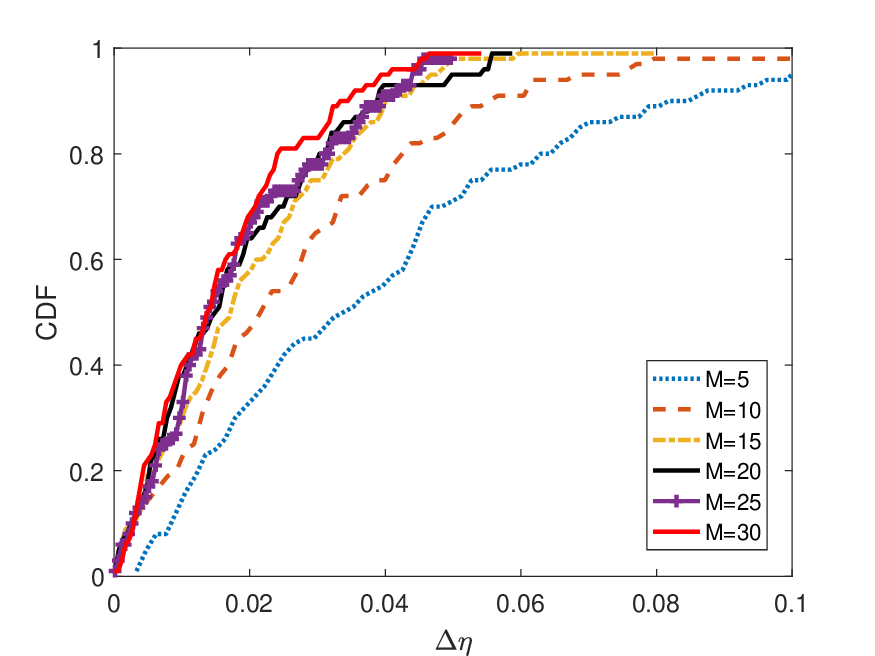}
\caption{The CDFs of the absolute estimation error in the case of different number of UEs.}
\label{fig4b}
\end{figure}

We now investigate the effect of the system size on the estimation of $\eta$. We consider an $M$-antenna MU-MIMO system with QPSK signaling.
Fig. \ref{fig4a} plots the CDFs of the absolute estimation error for different numbers of transmit antennas,
whilst Fig. \ref{fig4b} shows the results for different numbers of UEs.
In Fig. \ref{fig4a},  the number of UEs is $N = 20$. The numbers of antennas are 20, 32, 64, 128 and 256, respectively.
It is shown in Fig. \ref{fig4a} that the estimation performance improves as the increase of the number of transmit antennas.
In Fig. \ref{fig4b}, the numbers of UEs are 5, 10, 15, 20, 20, and 30, respectively. The number of transmit antennas is $M = 256$.
The results in Fig. \ref{fig4b} imply that one can obtain accurate estimates of $\eta$ with high probability for various numbers of
UEs. Besides, the estimation performance improves as the increase of the number of UEs. Interestingly, the estimation performance does
not improve distinctly when the number of UEs is above a certain threshold (e.g., $N\geq20$). We will revisit such a phenomenon with
a theoretical analysis in our future work.

%\subsubsection{Robustness in Different Error Conditions}
%
%Fig. \ref{fig3} presents the performance of the proposed EI-based moments matching method
% for a various values of $\eta$. In Fig. \ref{fig3}, we consider a massive MIMO system with 256 transmit antennas and 30 UEs.
%The results in Fig. \ref{fig3}
%indicate that accurate estimation of the noise parameter can be achieved with high probability in a wide range of $\eta$,
%which demonstrates the robustness of the proposed algorithm.
%
%
%\begin{figure}[!ht]
%\centering
%\includegraphics[width=0.48\textwidth]{figures/dif-eta}
%\caption{The CDFs of the absolute estimation error in a various values of $\eta$.}
%\label{fig3}
%\end{figure}

\subsection{The Accuracy of Channel Matrix Reconstruction}

\begin{figure}[!ht]
\centering
\includegraphics[width=0.48\textwidth]{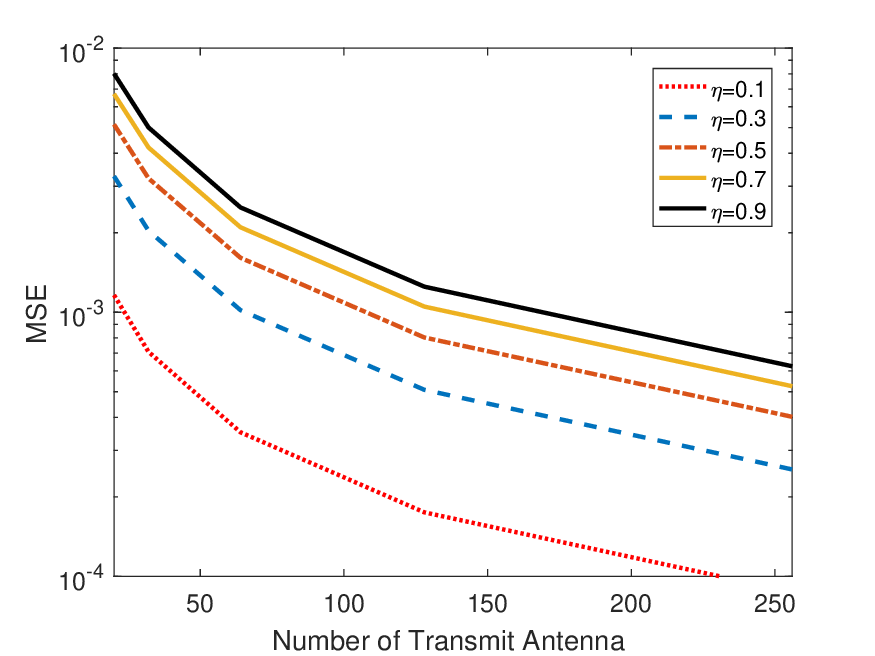}
\caption{MSE of channel matrix reconstruction with different number of transmit antennas.}
\label{fig6}
\end{figure}

We consider the case with $N=20$ UEs and $M=20,32,64,128,256$ transmit antennas. In this case, the CSI is reconstructed by the
proposed EI-based rotation invariant estimation method as shown in Algorithm 1.
Let ${\bf{\hat H}}$ be the estimate of ${\bf{H}}$.
Fig. \ref{fig6} illustrates the the mean square error (MSE) of channel matrix reconstruction, which is denoted by
\[{\rm{MSE = }}\frac{1}{{MN}}\left\| {{\bf{H}} - {\bf{\hat H}}} \right\|_{\mathop{\rm F}\nolimits} ^2.\]

It is shown in Fig. \ref{fig6} that the reconstruction error decreases as the increase in the number of transmit antennas.
Besides, the proposed method can provide accurate estimates of CSI from noisy CSI in
both small perturbation condition (e.g., $\eta=0.1$) and large perturbation condition (e.g., $\eta=0.9$),
indicating the robustness of the new method.
%the number transmit antennas is one of key parameters in the process of channel matrix reconstruction. In
%  massive MIMO systems are presumed to operate.

\subsection{BER Performance}

In this experiment, we investigate the BER performance of coarsely quantized MU-MIMO systems with imperfect CSI in three cases.
The BS is equipped with 256 transmit antennas and serves 30 UEs. For notation succinctness,
we use EI to denote the proposed eigen-inference precoder. WF and WF0 are respectively referred to
the WF precoding scheme implemented in the
case of perfect CSI and imperfect CSI (without reconstruction). We have similar definition for other compared precoding schemes.

\subsubsection{The Effect of Quantization Levels}

\begin{figure}[!ht]
\centering
\includegraphics[width=0.48\textwidth]{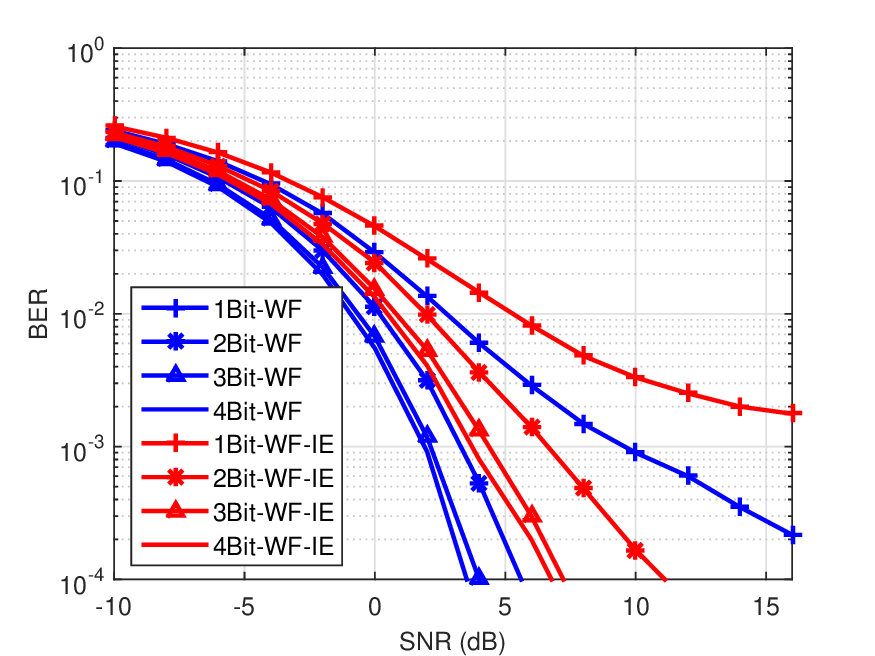}
\caption{Performance of the proposed precoder with different quantization levels
in the case of large MU-MIMO system with QPSK signaling.}
\label{fig7}
\end{figure}

Fig. \ref{fig7} shows the BER of the WF variants versus signal-to-noise ratios (SNRs) for different quantization levels (B-bit DACs).
The CSI-related parameter is fixed at $\eta=0.3$.
From Fig. \ref{fig7}, we observe that the performance of the WF precoder (with perfect CSI) and
the proposed WF-IE precoder (with imperfect CSI) improve with the increase of the quantization level.
The results  indicate that the proposed precoder can provide reliable
transmission of QPSK signaling in the case of imperfect CSI.
Specially, for $\{2,3,4 \}$-bit quantization, the performance gap to ideal BER (perfect CSI) with
that of WF-IE is only about 3 dB for a target BER of $10^{-3}$.
%In what follows, we will use 3-bit DACs.

\subsubsection{The Effect of the SNR}

\begin{figure}[!ht]
\centering
\includegraphics[width=0.48\textwidth]{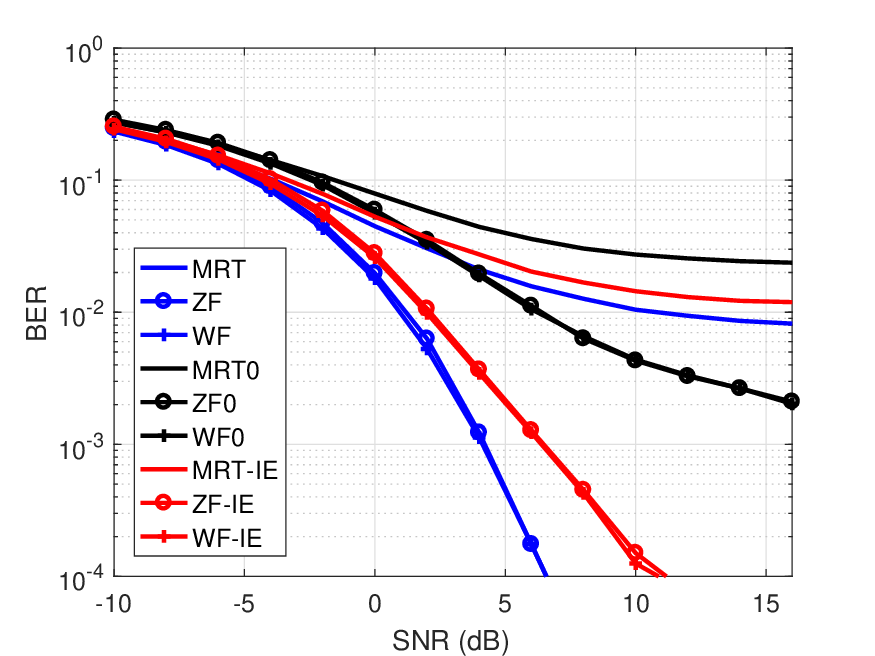}
\caption{Uncoded BER of the compared precoders in the case of a quantized massive MU-MIMO system (128 BS antennas and 20 UTs) with QPSK signaling.
The CSI-related parameter is fixed at $\eta=0.3$.}
\label{fig10}
\end{figure}

In Fig. \ref{fig10}, we consider a massive MIMO system with 4-bit DACs. It is shown
that the WF-IE percoder has similar performance
with the ZF-IE precoder and outperforms the MRT-IE precoder.
Moreover, the performance gap between the WF precoder with perfect CSI and the proposed WF-IE precoder (with imperfect CSI)
is remarkably smaller than that between the WF precoder and the WF0 precoder.
For example, for the same BER target of $10^{-3}$, the former is about 3 dB while the latter is more than 15 dB.
The result implies that the implementation of robust precoding algorithms for
coarsely quantized massive MIMO systems with imperfect
CSI is helpful and possible by employing proper signal
processing techniques (e.g., the proposed EI precoder).

\subsubsection{The Effect of Different Values of $\eta$}

\begin{figure}[!ht]
\centering
\includegraphics[width=0.48\textwidth]{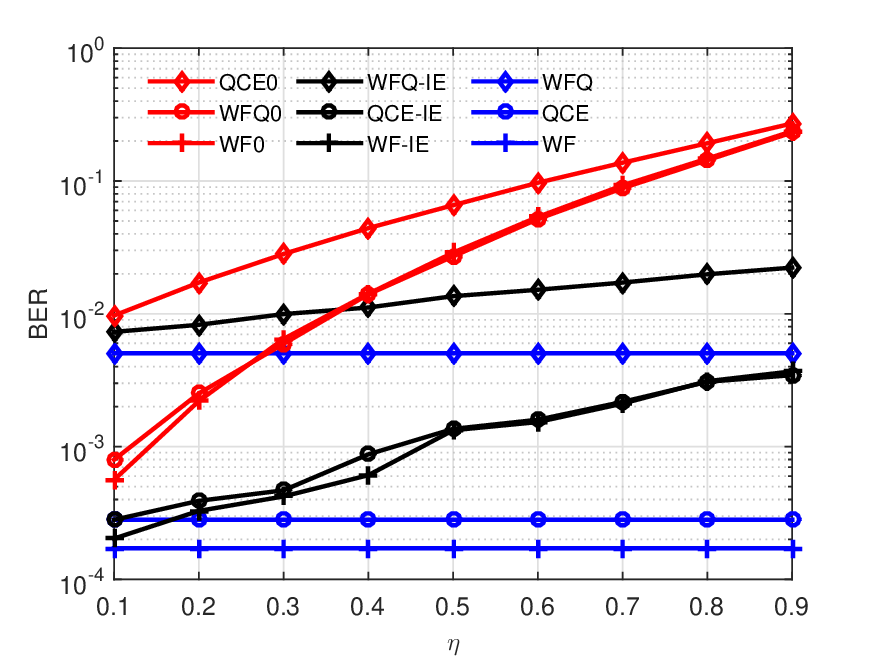}
\caption{Uncoded BER of the compared precoders for the MU-MIMO system with 4-bit DACs and different values of $\eta$.}
\label{fig9}
\end{figure}

In Fig. \ref{fig9}, we show the BER performance of all the precoders with different values of $\eta$. The SNR is fixed at 5 dB,
a low-to-moderate (typical) SNR.
In the case of perfect CSI, the WF percoder outperforms the WFQ precoder, which is in line with the result in \cite{Jacobsson2016Quantized}.
In the presence of imperfect CSI, it can be seen from Fig. \ref{fig9} that each precoder degrades as the increase of $\eta$.
However, they have different sensitivity, e.g., WFQ0, WF0 and QCE0 are more sensitive to errors in CSI than their counterparts
using the proposed CSI reconstruction method.
As a consequence, WFQ-IE, WF-IE and QCE-IE significantly outperform WFQ0, WF0 and QCE0, and the advantage increases as $\eta$ increases.
For example, WFQ0 and WF0 can approach a target BER of $10^{-3}$ only when $\eta < 0.15$, while WFQ-IE and WF-IE can achieve a target BER of
$10^{-3}$ for any $\eta < 0.45$.
Even in the limit case of $\eta = 0.9$, the proposed precoder can still achieve a BER of about $10^{-2.5}$.
On the whole, the results imply that the employment of simplified hardware (low-resolution DACs) could be possible without severe
BER performance loss in the case of imperfect CSI.

%\subsubsection{Comparison with the Nonlinear Precoders}

%WFQ0, WF0 and QCE0 are highly depend on the available CSI. WFQ0, WF0 and QCE0
%approaches a target BER of $10^{-3}$when  $\eta > 0.15$.
%On the other hand, WFQ-IE, WF-IE and QCE-IE are
%less sensitive due to the proposed CSI reconstruction methods. The proposed WF-IE precoder can obtain a target BER of $10^{-3}$ for any
%$\eta < 0.45$. In the limit case when $\eta = 0.9$, the proposed precoder can still achieve a BER of about $10^{-2.5}$.
%These results imply that the employment of
%simplified hardware (low-resolution DACs) for MU-MIMO with partial CSI could be possible without having severe
%BER performance degradation.

%
%So far, we have examined the performance of MIMO systems with QPSK signaling in the case of imperfect CSI.
%Here, we investigate the proposed precoder for different modulation schemes,
%QPSK, 8PSK and 16QAM. The CSI-related parameter is fixed as $\eta=0.3$.
%Fig. \ref{fig8} presents the performance of the proposed precoder compared to the
%WF precoder with perfect CSI in the three modulation modes.
%The results in Fig. \ref{fig8} indicate that the proposed precoder can provide reliable
%transmission of QPSK signaling in the case of imperfect CSI. The performance gap to ideal BER (perfect CSI) with WF-IE is only about 3 dB for a target
%BER of $10^{-3}$. For
%
%\begin{figure}[!ht]
%\centering
%\includegraphics[width=0.48\textwidth]{figures/ber-mode}
%\caption{Performance of the proposed precoder for a large MU-MIMO system with 4-bit DACs and different modulation modes.}
%\label{fig8}
%\end{figure}

\section{Conclusions}
\label{Sec:f}

In the paper, we developed a novel precoding scheme for coarsely quantized massive MU-MIMO systems in the presence of imperfect CSI.
Firstly, we provided some analysis using the block random matrix theory, based on which the limiting spectra distribution connection
between the true BSCA and noisy BSCA has been established. Then, with the obtained theoretical results, we proposed an EI-based moments
matching method to estimate the CSI-related noise level and a rotation invariant estimation method to reconstruct the CSI.
Finally, experimental results have demonstrated that, the proposed EI precoding scheme can significantly mitigate the deterioration
caused by imperfect CSI in coarsely quantized massive MU-MIMO systems.

%In our future work, we will focus on investigating more channel models, i.e., correlated Rayleigh and Rician fading \cite{Le2017Distributions}.

%We end this part by pointing out that there's plenty of room for performance improvement, i.e., investigating the more complex noise models and providing
%theoretical analysis for the phenomenon shown in Section \ref{ess}.

\section{Acknowledgements}
\label{Sec:g}

The authors would like to thank the AE and anonymous reviewers for their valuable suggestions and criticism which improve the quality of this work.

\begin{appendices}

\section{Proof of Lemma \ref{lma3}}
\label{APX1a}
%The technical results stated in Lemma \ref{lma1}, Lemma \ref{lma2}, and Lemma \ref{lma3}

$Proof:$ Following \cite{Tulino2004Random}, one can expand the $Stieltjes$ transform of a random matrix $\bf{A}$ as
\begin{equation}
\renewcommand\theequation{A.1}
\label{eqap11}
{{\mathfrak{g}}_{\bf{A}}}\left( z \right) =  - \frac{1}{z}\sum\limits_{k = 0}^\infty  {\frac{{\mathbb{E}\left[ {{{\bf{A}}^k}} \right]}}{{{z^k}}}}.
\end{equation}
If $\bf{{A}}$ is a square random matrix, i.e., ${\bf{A}} \in {\mathbb{C}}^{N \times N}$, then one can have, in large dimensional regime
\footnote{In large dimensional regime, the limiting spectra of $\bf{{A}}$ is self-averaging \cite{Couillet2012Robust, Tao2012Topics},
that is to say, the distribution of $\bf{{A}}$ can be regarded as the averaged empirical eigenvalue distribution of itself.},
\begin{equation}
\renewcommand\theequation{A.2}
\label{eqap12}
{\mathbb{E}\left[ {{{\bf{A}}^k}} \right]} \equiv \mathbb{E}\left[ {\frac{1}{N}{\rm tr}\left\{ {{{\bf{A}}^k}} \right\}} \right].
\end{equation}
The $Stieltjes$ transform of $\bf{D}$ can be written as
\begin{equation}
\renewcommand\theequation{A.3} \label{eqap13}
\begin{array}{c}
{{\mathfrak{g}}_{\bf{D}}}\left( z \right) =  - \frac{1}{z}\sum\limits_{k = 0}^\infty  {\frac{\mathbb{E}{\left[ {{{\bf{D}}^k}} \right]}}{{{z^k}}}}
 \equiv \frac{1}{{M + N}}\mathbb{E}\left[ {{\mathop{\rm tr}\nolimits} \left\{ {{{\left( {z{\bf{I}} - {\bf{D}}} \right)}^{ - 1}}} \right\}} \right].
\end{array}
\end{equation}
From the definition of $\bf D$ in \eqref{eqa6}:
\[{\bf{D}} = {\bf{B}}{{\bf{B}}^{\rm H}} = \left[ {\begin{array}{*{20}{c}}
{{\bf{H}}{{\bf{H}}^{\rm H}}}&0\\
0&{{{\bf{H}}^{\rm H}}{\bf{H}}}
\end{array}} \right],\]
one can have, using the inverse of the block matrix,
\begin{align}
\label{eqap14}
{{\mathfrak{g}}_{\bf{D}}}\left( z \right) &= \frac{1}{{M + N}}{\mathop{\rm tr}\nolimits} \left[ {\begin{array}{*{20}{c}}
{{{\left( {z{\bf{I}} - {\bf{H}}{{\bf{H}}^{\rm H}}} \right)}^{ - 1}}}&0 \notag\\
0&{{{\left( {z{\bf{I}} - {{\bf{H}}^{\rm H}}{\bf{H}}} \right)}^{ - 1}}}
\end{array}} \right] \notag\\
 &= \frac{1}{{M + N}}\left[ {{\mathop{\rm tr}\nolimits} {{\left( {z{\bf{I}} - {\bf{H}}{{\bf{H}}^{\rm H}}} \right)}^{ - 1}} + {\mathop{\rm tr}\nolimits}
 {{\left( {z{\bf{I}} - {{\bf{H}}^{\rm H}}{\bf{H}}} \right)}^{ - 1}}} \right] \notag\\
 &= \frac{M}{{M + N}}{{\mathfrak{g}}_{{\bf{H}}{{\bf{H}}^{\rm H}}}}\left( z \right) + \frac{N}{{M + N}}{{\mathfrak{g}}_{{{\bf{H}}^{\rm H}}{\bf{H}}}}
 \left( z \right). \tag{A.4}
\end{align}
From Lemma 3.1 in \cite{Couillet2012Random}, we have
\begin{equation}
\renewcommand\theequation{A.5} \label{eqap15}
{{\mathfrak{g}}_{{{\bf{H}}^{\rm H}}{\bf{H}}}}\left( z \right) = {\frac{M}{N}{{\mathfrak{g}}_{{\bf{H}}{{\bf{H}}^{\rm H}}}}\left( z \right) +
\frac{{M - N}}{N}\frac{1}{z}}.
\end{equation}
Let $q=M/N$, plugging (A.5) into \eqref{eqap14} gives
\begin{equation}
\renewcommand\theequation{A.6} \label{eqap16}
{{\mathfrak{g}}_{\bf{D}}}\left( z \right) = \frac{{2q}}{{q + 1}}{{\mathfrak{g}}_{{\bf{H}}{{\bf{H}}^{\rm H}}}}\left( z \right) +
\frac{{q - 1}}{{q + 1}}\frac{1}{z},
\end{equation}
which is the result shown in Lemma \ref{lma3}.

\section{Proof of Lemma \ref{lma1}}
\label{APX1b}
% The corresponding densities functions are respectively denoted by ${f_{\bf{B}}}\left( x \right)$ and ${f_{\bf{D}}}\left( x \right)$.
From \eqref{eqa6}, we have ${\bf{D}} = {\bf{B}}^2$. Let ${F_{\bf{B}}}\left( x \right)$ and ${F_{\bf{D}}}\left( x \right)$ be the empirical
eigenvalue distribution of ${\bf{B}}$ and ${\bf{D}}$, respectively.  Since the eigenvalues of ${\bf{B}}$ are the radicals of the eigenvalues
of ${\bf{D}}$, let $\left[ {\left( { - \sqrt b , - \sqrt a } \right) \cup \left( {\sqrt a ,\sqrt b } \right)} \right]$ be the support of
 the eigenvalues of ${\bf{B}}$ and recall the definition of the $Stieltjes$ transform in \eqref{eqa2}, then we have
\begin{align}
\label{eqap21}
{{\mathfrak{g}}_{\bf{B}}}\left( z \right) &= \int_{\sqrt a }^{\sqrt b } {\frac{1}{{z - x}}{\mathop{\rm d}\nolimits} {F_{\bf{B}}}\left( x \right) +
\int_{ - \sqrt b }^{ - \sqrt a } {\frac{1}{{z - x}}{\mathop{\rm d}\nolimits} {F_{\bf{B}}}\left( x \right)} } \notag \\
 &= \int_{\sqrt a }^{\sqrt b } {\frac{1}{{z - x}}{\mathop{\rm d}\nolimits} {F_{\bf{B}}}\left( x \right) + \int_{\sqrt a }^{\sqrt b } {\frac{1}{{z + x}}
 {\mathop{\rm d}\nolimits} {F_{\bf{B}}}\left( x \right)} } \notag \\
 &= \int_{\sqrt a }^{\sqrt b } {\frac{{2z}}{{{z^2} - {x^2}}}{\mathop{\rm d}\nolimits} {F_{\bf{B}}}\left( x \right)} \notag \\
 &= \int_a^b {\frac{z}{{{z^2} - y}}{\mathop{\rm d}\nolimits} {F_{{{\bf{B}}^2}}}\left( y \right)} \notag \\
 &= z{{\mathfrak{g}}_{\bf{D}}}\left( {{z^2}} \right) \tag{B.1} ,
\end{align}
which completes the proof of Lemma \ref{lma1}.

\section{Proof of Lemma \ref{lma2}}
\label{APX1c}

We start with the definition of the $\mathcal{M}$ transform \cite{Tulino2004Random} of a random matrix $\bf A$, which is denoted by
\begin{equation}
\renewcommand\theequation{C.1}
\label{eqap31}
{\mathcal{M}_{\bf{A}}}\left( z \right) = \sum\limits_{k = 1}^\infty  {\mathbb{E}\left[ {{{\bf{A}}^k}} \right]{z^k}},
\end{equation}
or
\begin{equation}
\renewcommand\theequation{C.2}
\label{eqap32}
{\mathcal{M}_{\bf{A}}}\left( z \right) = {z^{ - 1}}{{\mathfrak{g}}_{\bf{A}}}\left( {{z^{ - 1}}} \right) + 1.
\end{equation}
Besides, the $S$ transform can be expressed with the $\mathcal{M}$ transform as
\begin{equation}
\renewcommand\theequation{C.3}
\label{eqap33}
{S_{\bf{A}}}\left( z \right) = \frac{{1 + z}}{z}{\mathcal{M}}_{\bf{A}}^{ - 1}\left( z \right).
\end{equation}
From \eqref{eqthla1} (or (B.1)), it follows that
\begin{equation}
\renewcommand\theequation{C.4}
\label{eqap34}
{{\mathcal{M}}_{\bf{B}}}\left( z \right) = {z^{ - 1}}{{\mathfrak{g}}_{\bf{B}}}\left( {{z^{ - 1}}} \right) + 1 = {z^{ - 2}}{{\mathfrak{g}}_{\bf{D}}}
\left( {{z^{ - 2}}} \right) + 1 = {{\mathcal{M}}_{\bf{D}}}\left( {{z^2}} \right).
\end{equation}
Substituting (C.4) into (C.3) yields
\begin{align}
{S_{\bf{B}}}\left( z \right) &= \frac{{1 + z}}{z}{\mathcal{M}}_{\bf{B}}^{ - 1}\left( z \right) \notag \\
 &= \frac{{1 + z}}{z}{\left[ {{\mathcal{M}}_{\bf{D}}^{ - 1}\left( z \right)} \right]^{1/2}} \notag \\
 &= \frac{{1 + z}}{z}{\left[ {\frac{z}{{1 + z}}{S_{\bf{D}}}\left( z \right)} \right]^{1/2}} \notag,
\end{align}
or equivalently,
\begin{equation}
\renewcommand\theequation{C.5}
\label{eqap35}{\left[ {{S_{\bf{B}}}\left( z \right)} \right]^2} = \frac{{1 + z}}{z}{S_{\bf{D}}}\left( z \right),
\end{equation}
which completes the proof of Lemma \ref{lma2}.

\section{Proof of Theorem \ref{thmb1}}
\label{APX2}
With the technical results in Lemma \ref{lma3}, Lemma \ref{lma1} and Lemma \ref{lma2}, one can easily obtain
 Theorem \ref{thmb1} by means of the transforms of random matrices.

From \cite{Tulino2004Random}, the  $Stieltjes$ transform of ${{\bf{H}}{{\bf{H}}^{\rm H}}}$ is
\begin{equation}
\renewcommand\theequation{D.1} \label{eqap41}
{\mathfrak{g}_{{\bf{H}}{{\bf{H}}^{\rm H}}}}\left( z \right) = \frac{{1 - q - z - \sqrt {{z^2} - 2\left( {q + 1} \right)z +
{{\left( {q - 1} \right)}^2}} }}{{2qz}}.
\end{equation}
Plugging (D.1) into \eqref{eqthma3} (or (A.6)), we have
\begin{align}
\label{eqap42}
{{\mathfrak{g}}_{\bf{D}}}\left( z \right) &= \frac{{2q}}{{q + 1}}{{\mathfrak{g}}_{{\bf{H}}{{\bf{H}}^{\rm H}}}}\left( z \right) +
\frac{{q - 1}}{{q + 1}}\frac{1}{z} \notag \\
&= \frac{{ - z - \sqrt {{z^2} - 2\left( {q + 1} \right)z + {{\left( {q - 1} \right)}^2}} }}{{\left( {q + 1} \right)z}}. \tag{D.2}
\end{align}
With Lemma \ref{lma1}, one can obtain the $Stieltjes$ transform of $\bf{B}$, which is denoted by
\begin{equation}
\renewcommand\theequation{D.3} \label{eqap43}
{{\mathfrak{g}}_{\bf{B}}}\left( z \right) = \frac{{ - z^2 - \sqrt {{z^4} - 2\left( {q + 1} \right)z^2 + {{\left( {q - 1} \right)}^2}} }}
{{\left( {q + 1} \right)z}}.
\end{equation}
Given ${{\mathfrak{g}}_{\bf{B}}}\left( z \right)$, the inversion formula \cite{Tulino2004Random, Couillet2012Random}
that yields the limiting probability density function of the eigenvalues of ${\bf{B}}$ is
\begin{align}
\label{eqap44}
{{\mathfrak{g}}_{\bf{B}}}\left( z \right) &= \mathop {\lim }\limits_{y \to {0^ + }} \frac{1}{\pi }{\mathop{\Im}\nolimits}
\left[ {{g_{\bf{B}}}\left( {x + {\mathop{\rm i}\nolimits} y} \right)} \right] \notag \\
& = \frac{{\sqrt {\left( {{b^2} - {x^2}} \right)\left( {{x^2} - {a^2}} \right)} }}{{\left( {q + 1} \right)\pi \left| x \right|}}
 + \frac{{1 - q}}{{1 + q}}\delta \left( x \right). \tag{D.4}
\end{align}
where $a \le \left| x \right| \le b$ and $a = 1 - \sqrt q ,b = 1 + \sqrt q .$ This completes the proof of Theorem \ref{thmb1}.

\section{Proof of Theorem \ref{thmb2}}
\label{APX3}

For the BSCA ${\bf{\widetilde B}}$, from (D.3), we have
\begin{align}
{{\mathfrak{g}}_{\bf{B}}}\left( z \right) &= \frac{{ - z^2 - \sqrt {{z^4} - 2\left( {q + 1} \right)z^2 +
{{\left( {q - 1} \right)}^2}} }}{{\left( {q + 1} \right)z}}, \notag \\
&=\frac{{2q}}{{q + 1}}{{\mathfrak{g}}_{\bf{\mathord{\buildrel{\lower3pt\hbox{$\scriptscriptstyle\smile$}}
\over B} }}}\left( z \right) + \frac{{q - 1}}{{2qz}}, \tag{E.1}
\end{align}
where ${\bf{\mathord{\buildrel{\lower3pt\hbox{$\scriptscriptstyle\smile$}}
\over B} }}$ is an auxiliary matrix whose $Stieltjes$ transform satisfies (E.1). With Lemma \ref{lma3}, Lemma \ref{lma1}, we can have
\begin{equation}
\renewcommand\theequation{E.2} \label{eqap52}
{{\mathfrak{g}}_{\bf{\mathord{\buildrel{\lower3pt\hbox{$\scriptscriptstyle\smile$}}
\over B} }}}\left( z \right) = z{{\mathfrak{g}}_{{\bf H}{{\bf H}^{\rm H}}}}\left( {{z^2}} \right).
\end{equation}
Using (E.2) and Lemma \ref{lma2}, one can obtain the $S$ transform of ${\bf{\mathord{\buildrel{\lower3pt\hbox{$\scriptscriptstyle\smile$}}
\over B} }}$, which can be denoted by
\begin{equation}
\renewcommand\theequation{E.3}
\label{eqap53}{\left[ {{S_{\bf{\mathord{\buildrel{\lower3pt\hbox{$\scriptscriptstyle\smile$}}
\over B} }}}\left( z \right)} \right]^2} = \frac{{1 + z}}{z}{S_{{\bf H}{{\bf H}^{\rm H}}}}\left( z \right).
\end{equation}
The relation between the $R$ transform and the $S$ transform has to satisfy \cite{Couillet2012Random, Qiu2017Smart}
\begin{equation}
\renewcommand\theequation{E.4}
\label{eqap54}
\frac{1}{{R\left( z \right)}} = S\left( z{R\left( z \right)} \right).
\end{equation}
Plugging (E.3) into (E.4) gives
\begin{equation}
\renewcommand\theequation{E.5}
\label{eqap55}
{\left[ {\frac{1}{{{R_{{\bf{\mathord{\buildrel{\lower3pt\hbox{$\scriptscriptstyle\smile$}}
\over B} }}}}\left( z \right)}}} \right]^2} = {\left[ {{S_{{\bf{\mathord{\buildrel{\lower3pt\hbox{$\scriptscriptstyle\smile$}}
\over B} }}}}\left( {z{R_{{\bf{\mathord{\buildrel{\lower3pt\hbox{$\scriptscriptstyle\smile$}}
\over B} }}}}\left( z \right)} \right)} \right]^2} = \frac{{z + 1}}{z}{S_{{\bf{H}}{{\bf{H}}^{\rm H}}}}
\left( {z{R_{{\bf{\mathord{\buildrel{\lower3pt\hbox{$\scriptscriptstyle\smile$}}
\over B} }}}}\left( z \right)} \right)
\end{equation}
Since ${S_{{\bf{H}}{{\bf{H}}^{\rm H}}}}\left(z\right)=\frac{1}{{1 + qz}}$  \cite{Couillet2012Random, Tulino2004Random},
one can obtain the $R$ transform of
${\bf{\mathord{\buildrel{\lower3pt\hbox{$\scriptscriptstyle\smile$}}
\over B} }}$ by  solving the equation (E.5), which yields
\begin{equation}
\renewcommand\theequation{E.6}
\label{eqap56}{R_{{\bf{\mathord{\buildrel{\lower3pt\hbox{$\scriptscriptstyle\smile$}}
\over B} }}}}\left( z \right) = \frac{{ - 1 + q{z^2} + \sqrt {1 + 4{z^2} - 2q{z^2} + {q^2}{z^4}} }}{{2z}}.
\end{equation}
For the noisy BSCA ${\bf{\widetilde B}}$, one can similarly define an auxiliary matrix
${\bf{\mathord{\buildrel{\lower3pt\hbox{$\scriptscriptstyle\frown$}}
\over B} }}$ that
\begin{equation}
\renewcommand\theequation{E.7} \label{eqap57}
{{\mathfrak{g}}_{\bf{\mathord{\buildrel{\lower3pt\hbox{$\scriptscriptstyle\frown$}}
\over B} }}}\left( z \right) = z{{\mathfrak{g}}_{{\bf {\widetilde H}}{{\bf {\widetilde H}}^{\rm H}}}}\left( {{z^2}} \right),
\end{equation}

Combined with the addition law \cite{Tulino2004Random} shown in \eqref{eqa73} and Lemma 4.1 in \cite{Couillet2012Random},
the obtained $R$ transform of ${\bf{\mathord{\buildrel{\lower3pt\hbox{$\scriptscriptstyle\frown$}}
\over B} }}$ reads
\begin{equation}
\renewcommand\theequation{E.8}
\label{eqap58}{R_{{\bf{\mathord{\buildrel{\lower3pt\hbox{$\scriptscriptstyle\frown$}}
\over B} }}}}\left( z \right) = \frac{{ - 1 + q{z^2} + h\left( z \right) }}{{2z}},
\end{equation}
where
\begin{align}
h\left( z \right) &= \sqrt {1 + z^2\left( {4 + q\left( { - 2 + qz^2} \right)} \right)}   \notag \\
 &+ \sqrt {1 + {\alpha^2}z^2\left( {4 + q\left( { - 2 + z^2{\alpha^2}q} \right)} \right)}.  \notag
\end{align}
From \eqref{eqa3}, we can have
\begin{equation}
\renewcommand\theequation{E.9}
\label{eqap59}
{R_{{\bf{\mathord{\buildrel{\lower3pt\hbox{$\scriptscriptstyle\frown$}}
\over B} }}}}\left( {\mathfrak{g}_{{\bf{\mathord{\buildrel{\lower3pt\hbox{$\scriptscriptstyle\frown$}}
\over B} }}}} \left( z \right) \right) = z - \frac{1}{{\mathfrak{g}_{{\bf{\mathord{\buildrel{\lower3pt\hbox{$\scriptscriptstyle\frown$}}
\over B} }}}} \left( z \right)}.
\end{equation}
Finally,  plugging (E.7) and (E.8) into (E.9) and solving (E.9) yields the technical result shown in Theorem \ref{thmb2}, which completes the proof.

\end{appendices}

\bibliographystyle{IEEEtran}
\normalem
\bibliography{icc_leo}
\end{document}

%% file: preamble.tex
\usepackage{cite}

\usepackage{amssymb}
\usepackage{amsmath}
\usepackage{makeidx}
\usepackage{multicol}
\usepackage[abs]{overpic}
\usepackage{tabularx}
\usepackage{amsfonts}
\usepackage{array,color}
\usepackage{tabularx}
\usepackage{multirow}
\usepackage{graphics}
\usepackage{epsfig}
\usepackage{graphicx}
\usepackage{makeidx}
\usepackage{multicol}
\usepackage{amsthm}
\usepackage{bm}
\usepackage{booktabs}
\usepackage{ulem}
\usepackage{enumerate}
\usepackage[caption=false,font=footnotesize]{subfig}
\usepackage[font=small,labelsep=period]{caption}
\usepackage{algorithm}
\usepackage{algorithmic}
\usepackage{ulem}

\newtheorem{theorem}{theorem}[section]
\newtheorem{myTheo}{Theorem}[section]

\newtheorem{lemma}[theorem]{Lemma}

\newtheorem{proposition}[theorem]{Proposition}